\documentclass[amsmath,amssymb,twocolumn,aps]{revtex4}

\usepackage{epsfig}
\usepackage{graphicx}
\usepackage{color}

\begin{document}

\title{Ground state properties of sub-Ohmic spin-boson model with simultaneous diagonal and off-diagonal coupling}

\author{Nengji Zhou$^{1,2}$, Lipeng Chen$^{1}$, Yang Zhao$^{1}$, Dima Mozyrsky$^{3}$, Vladimir Chernyak$^{1,4}$, Yang Zhao$^{1}$\footnote{Electronic address:~\url{YZhao@ntu.edu.sg}}}
\date{\today}
\affiliation{$^1$Division of Materials Science, Nanyang Technological University, Singapore 639798, Singapore\\
$^2$Department of Physics, Hangzhou Normal University, Hangzhou 310046, China\\
$^3$Theoretical Division, Los Alamos National Laboratory, Los Alamos, New Mexico 87545, USA\\
$^4$Department of Chemistry, Wayne State University, Detroit, USA
}

\begin{abstract}
By employing the variational approach, density matrix renormalization group (DMRG), exact diagonalization as well as symmetry and mean-field analyses,
the ground state properties of the two-bath spin boson model with simultaneous diagonal and off-diagonal coupling are systematically studied
in the sub-Ohmic regime. A novel quantum phase transition from a doubly degenerate ``localized phase'' to
the other doubly degenerate ``delocalized phase'' is uncovered.  Via the multi-${\rm D}_1$ ansatz as the variational wave function,
transition points are determined accurately, consistent with the results from DMRG and exact diagonalization. An effective spatial dimension
$d_{\rm eff} = 2.37(6)$ is then estimated, which is found to be compatible with the mean-field prediction. Furthermore, the quantum phase transition is inferred to be of first order for the baths described by a continuous spectral density function. In the case of single mode, however, the transition is softened.
\end{abstract}
%\pacs{05.30.Jp, 03.65.Yz, 73.63.-b,}
%\keywords{special relativity; Finsler spacetime; projectively flat}

\maketitle

\section{Introduction}

As an archetype of open quantum systems, the spin-boson model \cite{Leggett, Weiss} finds a wide range of applications in condensed phase physics and physical chemistry in topics such as quantum computation \cite{qs1,qs2,qs3}, spin dynamics \cite{Leggett,dyna,dua}, biological molecules \cite{et1,et2} and quantum phase transition \cite{qpt1,qpt2,alv,win}. The spin-boson model consists of a two-level system coupled linearly to an environment bath represented by a set of harmonic oscillators. The coupling between the system and the environment can be characterized by a spectral function $J(\omega)$, which usually adopts a power law form in the low frequency regime $J(\omega)\propto\omega^s$. Depending on the value of $s$, there exist three distinct cases known as sub-Ohmic ($s<1$), Ohmic ($s=1$) and super-Ohmic ($s>1$) regimes. An interesting aspect of the spin-boson model concerns the quantum phase transition in the ground state. Recent theoretical studies \cite{qpt1,qpt2,alv,win} show that there is a second-order phase transition separating a non-degenerate delocalized phase from a doubly degenerate localized phase due to the competition between the tunneling and the environment induced dissipation in the sub-Ohmic regime. It is also well known that there exists a Kosterlitz-Thouless type phase transition in the Ohmic regime \cite{Leggett}.

The spin-boson model is similar to a one-exciton, two-site version of the Holstein model \cite{Holstein} widely used to study optical and transport properties of organic and biological molecules. In the Holstein model, the diagonal and off-diagonal exciton-phonon coupling are defined as non-trivial dependence of the exciton site energies and transfer integrals on the phonon coordinates, respectively \cite{Su}. Similarly, the diagonal and off-diagonal coupling in the spin-boson model denote bath-induced modulation of the spin bias and tunneling, respectively. Most studies on the quantum phase transition of the spin-boson model consider the coupling in the diagonal form, predominantly because identifying the quantum phase transition of the spin-boson model with simultaneous diagonal and off-diagonal coupling is a challenging problem from the theoretical point of view. Recent studies \cite{lv} utilized the Davydov ${\rm D}_1$ variational ansatz to investigate the quantum phase transition of the spin-boson model in the sub-Ohmic regime with the spin coupled diagonally and off-diagonally to a common bath. It is revealed that the off-diagonal coupling lifts the degeneracy in the localized phase, thereby removing the second-order phase transition. The interplay between the diagonal and off-diagonal coupling is thus known to give rise to a much richer phase diagram.

\begin{figure}[tbp]
  \centering
  \includegraphics[width=0.65\linewidth]{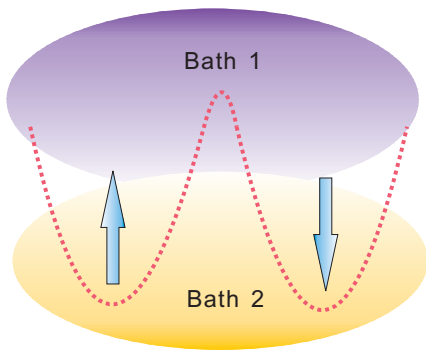}\\
  \caption{Schematics of the two-bath spin-boson model.
  }\label{fig0}
\end{figure}

To obtain a deeper insight into the competition between the diagonal and off-diagonal coupling, an additional phonon bath coupled to the spin off-diagonally can be taken into account, resulting in a two-bath spin-boson model (see Fig.~\ref{fig0}). This two-bath model is an appropriate low-energy description of a variety of physical systems, such as the excitonic energy transfer process in natural and artificial light-harvesting systems \cite{pach}, electromagnetic fluctuations of two linear circuits attached to a superconducting qubit \cite{you, card,raft}, two cavity fields coupled to a SQUID-based charge qubit \cite{liao}, and the process of thermal transport between two reservoirs coupled with a molecular junction \cite{ruok}.
In the case of zero bias and tunneling, the model exhibits a high level of symmetry, which can be described by a non-trivial central extension of the abelian symmetry group. The group theory analysis shows that the system's ground state is always doubly degenerate, and rendering invalid the picture of phase transition from degenerate to non-degenerate ground states. In other words, the ground state degeneracy does not necessarily support the spontaneous magnetization. Moreover, the quantum-to-classical correspondence fails in dealing with the two-bath model due to the sign problem \cite{guo}. Hence, it remains very challenging to understand the quantum phase transition of the two-bath model.

Previous numerical studies on the spin-boson model are typically based on the numerical renormalization group (NRG) \cite{Bulla,bulla2,Costi}, density matrix renormalization group (DMRG) \cite{guo}, the method of sparse polynomial space representation \cite{alv}, quantum Monte Carlo \cite{win}, the extended coherence state approach \cite{zyy, Wu, Chen} and the variational approach \cite{lv}. The results point to a second-order phase transition which is ascribed to the competition between the diagonal spin-bath coupling and the spin tunneling. Apart from the spin tunneling, the off-diagonal spin-bath coupling in the two-bath model can also provide a communication channel between spin-up and spin-down states.
It is thus interesting to investigate whether the quantum phase transition of the two-bath model retains its second order characteristics.

Motivated by these considerations, in this paper we aim to investigate the quantum phase transition of two-bath model in the simultaneous presence of the diagonal and off-diagonal coupling.
By using the variational approach, DMRG and exact diagonalization as well as symmetry and mean-field analyses, we conduct a comprehensive study on
the ground state properties of two-bath model, identify the picture of quantum phase transition and accurately determine transition points in the sub-Ohmic regime. A first-order quantum phase transition between the localized and delocalized states is inferred, and an effective spatial dimensional $d_{\rm eff}=2.37(6)$ is estimated, consistent with the mean-field prediction.
The paper is arranged as follows. In Sec.~II, the two-bath model is described, and the analyses based on symmetry and mean field are performed.
In Sec.~III and IV, the numerical results are presented for the quantum phase transition of the two bath model coupled to the baths described by the single mode and continuous spectral density function, respectively. Finally, the discussion and conclusions are presented in Sec.~V.

\section{Model and analysis}
\subsection{Model}
The standard Hamiltonian of the spin-boson model can be written as
\begin{eqnarray}\label{sbm1_h}
\hat{H}_{\textrm{SBM}}&=&\frac{\varepsilon}{2}\sigma_z-\frac{\Delta}{2}\sigma_x+\sum_l \omega_l b_l^\dag b_l\nonumber\\
&+&\frac{\sigma_z}{2}\sum_l \lambda_l(b^\dag_l+b_l),
\end{eqnarray}
where $\varepsilon$ is spin bias, $\sigma_{x}$ and $\sigma_z$ are pauli matrices, $\Delta$ is the tunneling constant,
$\omega_l$ denotes the frequency of the $l$-th effective bath mode for which $b_l$($b^{\dagger}_{l}$) represnts the phonon annihilation (creation) operator, and $\lambda_l$ signifies the coupling amplitude with the spin. The spectral density function is
\begin{equation}\label{dic_spec_func}
  J(\omega)=\sum_l \lambda^2_l \delta(\omega-\omega_l).
\end{equation}
Generally, it is convenient to rewrite Eq.~(\ref{sbm1_h}) into its continuous form
\begin{eqnarray}\label{sbm1_ctnu_h}
\hat{H} & = &\frac{\varepsilon}{2}\sigma_z-\frac{\Delta}{2}\sigma_x+\int^{\omega_c}_{0} g(\omega)b_{\omega}^\dag b_{\omega} \nonumber\\
&+&\frac{\sigma_z}{2}\int^{\omega_c}_{0}h(\omega)(b^\dag_{\omega}+b_{\omega}),
\end{eqnarray}
where $b_{\omega}$ and $b^\dagger_{\omega}$ are the continuous $b_{l}$ and $b^\dagger_{l}$, $g(\omega)$ is the dispersion relation, and $h(\omega)$ is the coupling function.
As indicated in Refs. \cite{chinmap} and \cite{Bulla}, $g(\omega)$ and $h(\omega)$ obey
\begin{equation}\label{gh}
\textrm{J}(\omega) = \pi\frac{dg^{-1}(\omega)}{d\omega}h^2(g^{-1}(\omega)),
\end{equation}
with $g^{-1}(\omega)$ being the inverse function of $g(\omega)$. A logarithmic discretization procedure
is adopted by dividing the phonon frequency interval $[0, \omega_c]$ into $M$ intervals
$[\Lambda^{-(l+1)},\Lambda^{-l}]\omega_c$ ($l=0,1,\ldots, M-1)$ and choosing $h(g^{-1}(\omega))$ as constant in each interval \cite{Bulla, zyy}. Where $M$ is the number of effective bath modes, and $\omega_c$ is the maximum frequency in the bath.
Then, the parameters $\omega_l$ and $\lambda_l$ in Eq.~(\ref{sbm1_h}) can be obtained by
\begin{eqnarray}
\lambda_l^2 & = & \int^{\Lambda^{-l}\omega_c}_{\Lambda^{-l-1}\omega_c}dxJ(x) \nonumber \\
\omega_l & = & \lambda^{-2}_l \int^{\Lambda^{-l}\omega_c}_{\Lambda^{-l-1}\omega_c}dxJ(x)x
\label{sbm1_dis}
\end{eqnarray}
Note that infinite bath modes are considered via the integration of the continuous spectral density $J(\omega)$,
although the number of effective bath modes $M$ is finite.

In this paper, we primarily aim to study the two-bath model, for which the Hamiltonian is written as
\begin{eqnarray}
\hat{H}_{\textrm {TBSBM}} & = & \frac{\varepsilon}{2}\sigma_z-\frac{\Delta}{2}\sigma_x + \sum_{l,i} \omega_l b_{l,i}^\dag b_{l,i} \nonumber \\
 & + & \frac{\sigma_z}{2}\sum_l \lambda_l(b^\dag_{l,1}+b_{l,1})  \nonumber \\
 & + & \frac{\sigma_x}{2}\sum_l \phi_l(b^\dag_{l,2}+b_{l,2}),
\label{Ohami}
\end{eqnarray}
where the subscript $i=1,2$ is introduced to distinguish the two baths, and $\lambda_l$ and $\phi_l$ are the diagonal and off-diagonal coupling amplitude, respectively, which determine spectral densities,
\begin{equation}\label{OspectraZ}
J_z(\omega)  =  2\alpha\omega_c^{1-s}\omega^s, \qquad J_x(\omega)  =  2\beta\omega_c^{1-\bar{s}}\omega^{\bar{s}}.
\end{equation}
Where $\alpha$ and $\beta$ are dimensionless coupling strengths, and the frequency cut off $\omega_c$ is set to be unity throughout this paper. The two boson baths are characterized by the spectral exponents $s$ and $\bar{s}$, accounting for the diagonal and off-diagonal, respectively.

\subsection{Symmetry arguments}

For nonzero values of $\alpha,\beta,\varepsilon$, and $\Delta$, the system Hamiltonian does not possess any symmetry. Therefore, in this work we assume that $\varepsilon=0$, and focus on the case of $\varepsilon=\Delta=0$ as it corresponds to scenarios with much stronger symmetry (the case of $\Delta=0$ can be reduced similarly by an obvious rotation).
We introduce the notation
\begin{eqnarray}
\label{inversion-phonons} P_{1}= \textrm{e}^{i\pi \sum_{n}b^{\dagger}_{n,1}b_{n,1}}, \;\;\; P_{2}= \textrm{e}^{i\pi\sum_{n}b^{\dagger}_{n,2}b_{n,2}},
\end{eqnarray}
and consider the operators
\begin{eqnarray}
\label{symmetry-generators} \mathcal{P}_{x}^{\pm}= \pm \sigma_{x}P_{1}, \;\;\; \mathcal{P}_{z}^{\pm}= \pm \sigma_{z}P_{2}
\end{eqnarray}
that act in the system space of states, and obviously commute with the system Hamiltonian when $\varepsilon=\Delta=0$. Taking the product of the above, we obtain
\begin{eqnarray}
\label{symmetry-generators-2} \mathcal{P}^{\gamma\zeta}= \mathcal{P}_{z}^{\gamma}\mathcal{P}_{x}^{\zeta} = i\gamma\zeta\sigma_{y}P_{1}P_{2},
\end{eqnarray}
where $\gamma,\zeta,\gamma\zeta=\pm$ clearly obey a product rule. A straightforward verification shows that eight operators ${\cal I}^{\pm},{\cal P}_{x}^{\pm},{\cal P}_{z}^{\pm},{\cal P}^{\pm}$, where ${\cal I}^{\pm}=\pm{\rm id}$, form a non-abelian group $G$, whose center (i.e., the set of elements that commute with any element of the group) is represented by $\{{\cal I}^{\pm}\}$. We thus have the factor group $G/\{{\cal I}^{\pm}\} \cong \mathbb{Z}_{2}\oplus \mathbb{Z}_{2}$ that is an abelian group. Stated differently, the non-abelian symmetry group $G$ of the two-bath model with zero bias and tunneling is given by a non-trivial central extension of the abelian group $\mathbb{Z}_{2}\oplus \mathbb{Z}_{2}$. The set of its unitary irreducible representations is given by four one-dimensional representations, characterized by the trivial action of the group center, and therefore labeled by four irreducible representations of the abelian group $\mathbb{Z}_{2}\oplus \mathbb{Z}_{2}$, characterized by a non-trivial action of the center, or more specifically the elements ${\cal I}^{\pm}$ are represented by the operators $\pm{\rm id}$. Since by definition the operators ${\cal I}^{\pm}$ act in the space of states as $\pm{\rm id}$, only the two-dimensional representation participates in the decomposition of the space of states in irreducible representations. Furthermore, by the Schur lemma, all energy levels, in particular the ground state, are necessarily doubly degenerate.

In the $\Delta \ne 0$ case, the symmetry is reduced to the abelian subgroup $G_{x}\subset G$ that consists of two elements $G_{x}= \{{\cal I}^{+},{\cal P}_{x}^{+}\}$, so that $G_{x}\cong \mathbb{Z}_{2}$, and the quantum phase transition occurs between the phase with spontaneous magnetization in the $z$-direction, characterized by a double-degenerate ground state, and a symmetric phase with no spontaneous magnetization $\langle\sigma_{z}\rangle$ and non-degenerate ground state. The above picture is quite similar to the phase transition in the standard spin-boson model with one diagonally-coupled bath. The situation in the $\Delta=0$ case is quite different. Firstly, due to symmetry considerations presented above, the system ground state is always doubly degenerate, and the phase transition from degenerate to non-degenerate ground state disappears. Secondly, one can, and should, consider spontaneous magnetization $\langle\sigma_{z}\rangle$ and $\langle\sigma_{x}\rangle$ in the $z$- and $x$-directions, respectively. In what follows we will first describe the symmetry-based picture of the phase transition in the $\Delta=0$ case, and then give further support to the presented scenario with numerical simulations, based on the variational and DMRG approaches.

We start with noting that, while non-degenerate ground state does not support spontaneous magnetization by mere symmetry arguments, either does the ground-state degeneracy necessarily. Rather, the latter merely creates an opportunity for spontaneous magnetization to occur. Indeed, to consider the dependence of spontaneous magnetization $\langle\sigma_{z}\rangle$ on bias $\varepsilon$, we switch on a very weak ``magnetic field,'' which in our case introduces a weak, yet non-zero bias $\varepsilon$.
For a non-degenerate ground state we will have $\langle\sigma_{z}\rangle \sim \varepsilon$, which corresponds to finite susceptibility. In the case of degenerate ground state the additional term $(\varepsilon/2)\sigma_{z}$ can eliminate the degeneracy, and we will obtain a finite value of $\langle\sigma_{z}\rangle$ for $\varepsilon \to 0$, given by the expectation value of the $\sigma_{z}$ operator evaluated with respect to the non-degenerate ground state. Therefore, to ascertain whether the symmetry is actually broken one needs to evaluate the projection of the $\sigma_{z}$ or $\sigma_{x}$ operator onto the two-dimensional subspace of the ground states. This can be done by invoking a convenient basis set of the eigenstates of ${\cal P}_{z}= {\cal P}_{z}^{+}$ or ${\cal P}_{x}= {\cal P}_{x}^{+}$ operator. Since ${\cal P}_{z}^{2}= {\cal I}$, the eigenvalues would be $\pm 1$. Let ${\cal P}_{z}|\psi\rangle = |\psi\rangle$; then ${\cal P}_{z}{\cal P}_{x}|\psi\rangle = -{\cal P}_{x}{\cal P}_{z}|\psi\rangle = -{\cal P}_{x}|\psi\rangle$, so that our basis is given by $(|\psi\rangle,{\cal P}_{x}|\psi\rangle)$. A straightforward computation yields $\langle\psi|{\cal P}_{x}\sigma_{z}{\cal P}_{x}|\psi\rangle= -\langle\psi|\sigma_{z}|\psi\rangle$ and $\langle\psi|{\cal P}_{x}\sigma_{z}|\psi\rangle= \langle\psi|\sigma_{z}{\cal P}_{x}|\psi\rangle= 0$, as well as $\langle\psi|{\cal P}_{x}\sigma_{x}{\cal P}_{x}|\psi\rangle= \langle\psi|\sigma_{x}|\psi\rangle=0$.  $\langle\psi|\sigma_{z}{\cal P}_{z}|\psi\rangle = \langle\psi|{\cal P}_{2}|\psi\rangle$  and $\langle\psi|\sigma_{x}{\cal P}_{x}|\psi\rangle = \langle\psi|P_{1}|\psi\rangle$ are also  derived. Denoting by $s_{x}$ and $s_{z}$ the operators, acting in the two-dimensional subspace of ground states, represented by the corresponding Pauli matrices in the basis set as introduced above, we arrive at
\begin{eqnarray}
\label{sigma-project} {\cal Q}\sigma_{z}= \langle\psi|P_{2}|\psi\rangle s_{z}, \;\;\; {\cal Q}\sigma_{x}= \langle\psi|P_{1}|\psi\rangle s_{x},
\end{eqnarray}
where ${\cal Q}$ denotes the projection onto the subspace of the ground states. It follows directly from Eq.~(\ref{sigma-project}) that the spontaneous magnetization
\begin{eqnarray}
\label{magnetization-overlap} |\langle\sigma_{z}\rangle|= \langle\psi|P_{2}|\psi\rangle, \;\;\; |\langle\sigma_{x}\rangle| = \langle\psi|P_{1}|\psi\rangle,
\end{eqnarray}
can be expressed in terms of overlaps of properly chosen system states. This means that although due to the symmetry the system ground state is always doubly degenerate, the symmetry can be broken or independent of whether the corresponding overlap in Eq.~(\ref{magnetization-overlap}) vanishes or not.

We are now in a position to lay out a picture of the phase transition, which will be verified by numerical simulations to come. For $\varepsilon=\Delta=0$ there is a phase transition that for given $\alpha$ occurs at $\beta= \beta_{{\rm c}}(\alpha)$, so that for $\beta < \beta_{{\rm c}}$ the system is in a phase with $\langle\sigma_{z}\rangle \ne 0$ and $\langle\sigma_{x}\rangle =0$, whereas for $\beta > \beta_{{\rm c}}$, the opposite trend ensues. This implies that one of the spontaneous magnetizations is always non-zero while the other necessarily vanishes.

\subsection{Mean field analysis}

The two-bath model can be treated using an approach presented in Appendix~A for the case of the standard spin-boson model counterpart. In this subsection, we present an alternate, completely equivalent approach to study two-bath model, in order to emphasize its connection with the theory of stochastic processes and the probability theory. Consider a Gaussian stochastic process for periodic $\bm{} \bm{B}(\tau+ \beta)= \bm{B}(\tau)$ trajectories $\bm{B}(\tau)= (B_{x}(\tau), B_{z}(\tau))$ with the probability measure
\begin{eqnarray}
\label{measure} d\mu(\bm{B})& = & e^{-S_{0}(\bm{B})}{\cal D}\bm{B}, \nonumber \\
S_{0}(\bm{B}) & = & \frac{1}{2\beta}\sum_{n}\left(\frac{|\tilde{B}_{x}(\omega_{n})^{2}|}{K_{x}(\omega_{n})}+ \frac{|\tilde{B}_{z}(\omega_{n})^{2}|}{K_{z}(\omega_{n})}\right),
\end{eqnarray}
where ${\cal D}$ is defined as a differential in the path integral, and
the time-ordered Matsubara Green functions of the spin operators adopt a form
{\footnotesize
\begin{eqnarray} \label{spin-corr-stoch}
& & \left\langle \sigma_{j_{1}}(\tau_{1})\ldots \sigma_{j_{s}}(\tau_{s}) \right\rangle_{H}   \nonumber \\
&=& Z^{-1}{\rm Tr}\left(T\left(\hat{\sigma}_{j_{1}}(\tau_{1})\ldots \hat{\sigma}_{j_{s}}(\tau_{s})\right)e^{-\beta H}\right) \nonumber \\
 &=& Z^{-1}\left\langle{\rm Tr}\left(T\left(\sigma_{j_{1}}(\tau_{1};\bm{B})\ldots \sigma_{j_{s}}(\tau_{s};\bm{B})\right) {\cal U}(\beta;\bm{B})\right)\right\rangle_{S_{0}(\bm{B})} \\
 &=& Z^{-1}\int d\mu(\bm{B}){\rm Tr}\left(T\left(\sigma_{j_{1}}(\tau_{1};\bm{B})\ldots \sigma_{j_{s}}(\tau_{s};\bm{B})\right) {\cal U}(\beta;\bm{B})\right) \nonumber \\
&=& Z^{-1}\int {\cal D}\bm{B}{\rm Tr}\left(T\left(\sigma_{j_{1}}(\tau_{1};\bm{B})\ldots \sigma_{j_{s}}(\tau_{s};\bm{B})\right) {\cal U}(\beta;\bm{B})\right)e^{-S_{0}(\bm{B})}.  \nonumber
\end{eqnarray}  }
In the equation above, we have denoted
{\footnotesize
\begin{eqnarray} \label{define-U}
{\cal U}(t;\bm{B}) & = & T\exp\left(-\int_{0}^{t}d\tau\left((B_{x}(\tau)+ \varepsilon)\sigma_{x}+ (B_{z}(\tau)+ \Delta)\sigma_{z}\right)\right), \nonumber \\
\sigma_{j}(\tau;\bm{B}) &  = &  {\cal U}(\tau;\bm{B})\sigma_{j}{\cal U}^{-1}(\tau;\bm{B}),
\end{eqnarray} }
The partition function is given as
{\small
\begin{eqnarray}
\label{Z-TBSB}
Z & = & {\rm Tr}e^{-\beta H} = \int d\mu(\bm{B}){\rm Tr}{\cal U}(\beta;\bm{B}) \nonumber \\
& = & \int {\cal D}\bm{B}{\rm Tr}{\cal U}(\beta;\bm{B})e^{-S_{0}(\bm{B})} = \int{\cal D}\bm{B}e^{-S_{{\rm eff}}(\bm{B})},
\end{eqnarray} }
and the form of the effective action is
\begin{eqnarray} \label{S-eff}
S_{{\rm eff}}(\bm{B})= S_{0}(\bm{B})- \ln{\rm Tr}{\cal U}(\beta;\bm{B}).
\end{eqnarray}
Note that Eq.~(\ref{spin-corr-stoch}) provides formal expressions for the Matsubara spin correlation functions in terms of stochastic averaging.
In the most interesting case of $\varepsilon= \Delta= 0$, Eq.~(\ref{define-U}) adopts a form
\begin{eqnarray}
\label{define-U-2} {\cal U}(t;\bm{B}) & = & T\exp\left(-\int_{0}^{t}d\tau\left(B_{x}(\tau)\sigma_{x}+ B_{z}(\tau)\sigma_{z}\right)\right) \nonumber  \\
& = & T\exp\left(-\int_{0}^{t}d\tau\bm{B}(\tau)\cdot \bm{\sigma}\right).
\end{eqnarray}

The above representation for the spin correlation functions, including the partition function, is readily obtained by introducing the collective coordinates $B_{x}= \sum_{k}g_{k}x_{k}$ and $B_{z}= \sum_{k}f_{k}z_{k}$, associated with the two uncorrelated baths, which fully describe the spin-bath coupling. Further, we note that the spin correlation functions in our quantum system are fully determined by the two-point correlation functions of the collective coordinates. These functions can thus be reproduced by a Gaussian stochastic model with the spin coupled to the stochastic variable $\bm{B}(\tau)$, provided the two-point correlation functions of the stochastic variables coincide with the quantum correlation functions of the corresponding collective-coordinate variables. The latter condition is satisfied if (and only if) the functions $K_{x}(\omega_{n})$ and $K_{z}(\bar{\omega}_{n})$ satisfy Eq.~(\ref{k}).

To obtain the mean-field picture of the phase transition, we consider the value of the effective action for $\bm{B}$ independent of $\tau$. In this case, we have
\begin{eqnarray}\label{E-eff}
S_{{\rm eff}}(\bm{B}) & = & \beta E_{{\rm eff}}(\bm{B}), \nonumber \\
E_{{\rm eff}}(\bm{B}) & = &  \frac{B_{x}^{2}}{2K_{x}(0)}+ \frac{B_{z}^{2}}{2K_{z}(0)}- \sqrt{B_{x}^{2}+ B_{z}^{2}},
\end{eqnarray}
where the contribution of the lower eigenvalue is neglected in the calculation of ${\rm Tr}({\cal U}(\beta;\bm{B}))$. Minimizing the effective energy with respect to $\bm{B}$, we obtain
\begin{eqnarray}
\label{E-eff-min} B_{x} &=& \pm K_{x}(0), \;\;\; B_{z}=0, \;\;\; {\rm for} \;\;\; K_{x}(0) > K_{z}(0), \nonumber \\ B_{x} &=& 0, \;\;\; B_{z}= \pm K_{z}(0), \;\;\; {\rm for} \;\;\; K_{x}(0) < K_{z}(0).
\end{eqnarray}
This will lead to a phase transition (on the mean field level of theory) at $K_{x}= K_{z}$. In each of the two phases, we have non-zero spontaneous magnetization in one direction, whereas there is no spontaneous magnetization in the other. Moreover, at the mean field level of theory, the phase transition is of first order, since there is spontaneous magnetization at the phase transition point.

Note that at the phase transition point the system has higher $U(1)$ symmetry, at least on the mean-field level. As we know in systems with local interactions, strong transverse fluctuations destroy spontaneous magnetization for the space dimension $d \le 2$. Although our model is one-dimensional, the nonlocal nature of interactions gives rise to an effective dimension that depends on the parameter $s$ and should be identified by considering the properties of long-range fluctuations around the mean-field solution. In what follows we argue that according to the aforementioned criterion, we have $d_{{\rm eff}}= 3-2s$, so that the critical value of $s$ is $s=1/2$, i.e., for $s< 1/2$ and $1/2 < s < 1$ the phase transition is first- and second-order, respectively.

\section{Single-mode baths}

\subsection{exact diagonalization results}

We first explore the ground state properties of a spin coupled to two single-mode baths using exact diagonalization. The corresponding Hamiltonian can be written as
\begin{eqnarray}\label{SBM_MODE}
H&=&\frac{\varepsilon}{2}\sigma_z-\frac{\Delta}{2}\sigma_x+\omega{(b_1^{\dagger}b_1+b_2^{\dagger}b_2)} \nonumber \\
&&+\frac{\sigma_z}{2}\lambda{(b_1^{\dagger}+b_1)}+\frac{\sigma_x}{2}\phi{(b_2^{\dagger}+b_2)},
\end{eqnarray}
where $\lambda$ and $\phi$ are diagonal and off-diagonal coupling constants, respectively. For convenience, we assume $|\psi_{\uparrow}\rangle$ and $|\psi_{\downarrow}\rangle$ are the bosonic states corresponding to the spin up and down states, which can be expanded in a series of Fock states $|k\rangle=[(b^{\dagger})^k/\sqrt{k!}]|0\rangle$ as follows
\begin{equation}\label{Expand1}
|\psi_{\uparrow}\rangle=\sum_{k_1k_2}^{N_{\rm tr}}c_{k_1k_2}|k_1k_2\rangle,
\end{equation}
\begin{equation}\label{Expand2}
|\psi_{\downarrow}\rangle=\sum_{k_1k_2}^{N_{\rm tr}}d_{k_1k_2}|k_1k_2\rangle,
\end{equation}
where $c_{k_1k_2}(d_{k_1k_2})$ are the coefficients with respect to a series of $\{k_1,k_2\}$ for the two bosonic baths with diagonal and off-diagonal coupling, respectively, and $N_{\rm tr}$ is the bosonic truncated number defined as a cutoff value of the phonon occupation number. In this work, $N_{\rm tr}=40$ is used in the exact diagonalization to label a truncated Hilbert space, sufficient large for the ground-state energy to converge.

The Schr\"{o}dinger equations of the Hamiltonian shown in Eq.~(\ref{SBM_MODE}) are then derived as
\begin{eqnarray}\label{EQ1}
\frac{\varepsilon}{2}|\psi_{\uparrow}\rangle+\omega{(b_1^{\dagger}b_1+b_2^{\dagger}b_2)}|\psi_{\uparrow}\rangle+\frac{\lambda}{2}{(b_1^{\dagger}+b_1)}|\psi_{\uparrow}\rangle
\nonumber \\
-\frac{\Delta}{2}|\psi_{\downarrow}\rangle+\frac{\phi}{2}{(b_2^{\dagger}+b_2)}|\psi_{\downarrow}\rangle = E|\psi_{\uparrow}\rangle,
\end{eqnarray}
\begin{eqnarray}\label{EQ2}
-\frac{\varepsilon}{2}|\psi_{\downarrow}\rangle+\omega{(b_1^{\dagger}b_1+b_2^{\dagger}b_2)}|\psi_{\downarrow}\rangle-\frac{\lambda}{2}{(b_1^{\dagger}+b_1)}|\psi_{\downarrow}\rangle
\nonumber \\
-\frac{\Delta}{2}|\psi_{\uparrow}\rangle+\frac{\phi}{2}{(b_2^{\dagger}+b_2)}|\psi_{\uparrow}\rangle = E|\psi_{\downarrow}\rangle.
\end{eqnarray}
After substituting Eqs.~(\ref{Expand1}) and (\ref{Expand2}) and left multiplying the bosonic states on both sides of Eqs.~(\ref{EQ1}) and (\ref{EQ2}), one has
{\small
\begin{eqnarray}\label{EOM}
&&\frac{\varepsilon}{2}c_{k_1k_2}+\omega{(k_1+k_2)}c_{k_1k_2}+\frac{\lambda}{2}(c_{k_1-1,k_2}\sqrt{k_1} + c_{k_1+1,k_2} \nonumber \\
&&\sqrt{k_1+1})-\frac{\Delta}{2}d_{k_1k_2} + \frac{\phi}{2}(d_{k_1,k_2-1}\sqrt{k_2}+d_{k_1,k_2+1}\sqrt{k_2+1}) \nonumber \\
&&=Ec_{k_1k_2}, \nonumber \\
&&-\frac{\varepsilon}{2}d_{k_1k_2}+\omega{(k_1+k_2)}d_{k_1k_2}-\frac{\lambda}{2}(d_{k_1-2,k_2}\sqrt{k_1}+ d_{k_1+1,k_2} \nonumber \\
&&\sqrt{k_1+1})-\frac{\Delta}{2}c_{k_1k_2} + \frac{\phi}{2}(c_{k_1,k_2-1}\sqrt{k_2}+c_{k_1,k_2+1}\sqrt{k_2+1})  \nonumber \\
&&=Ed_{k_1,k_2}.
\end{eqnarray}
}
The expectation value of $\sigma_z$ and $\sigma_x$ can be derived as
\begin{eqnarray}\label{Expectation}
\langle{\sigma_x}\rangle &=&\sum_{k_1k_2}c_{k_1k_2}^{*}d_{k_1k_2}+d_{k_1k_2}^{*}c_{k_1k_2}, \nonumber \\
\langle{\sigma_z}\rangle &=&\sum_{k_1k_2}|c_{k_1k_2}|^2-|d_{k_1k_2}|^2.
\end{eqnarray}
The von Neumann entropy $S_{\rm v-N}$, also known as the entanglement entropy\cite{Bennett,Zhao}, that characterizes the entanglement between the spin and the surrounding bath is also introduced \cite{Costi,Amico},
\begin{equation} \label{Entropy}
S_{\textrm{v-N}}=-\omega_{+}\log\omega_{+}-\omega_{-}\log\omega_{-},
\end{equation}
where $\omega_{\pm}=(1\pm\sqrt{\langle{\sigma_x}\rangle^2+\langle{\sigma_y}\rangle^2+\langle{\sigma_z}\rangle^2})/2$.
It should be noted that $\langle{\sigma_y}\rangle \equiv 0$ due to Hamiltonian invariance under the transformation $\sigma_y\rightarrow{-\sigma_y}$ \cite{Costi}.

\begin{figure}[tbp]
\centering
\includegraphics*[width=0.7\linewidth]{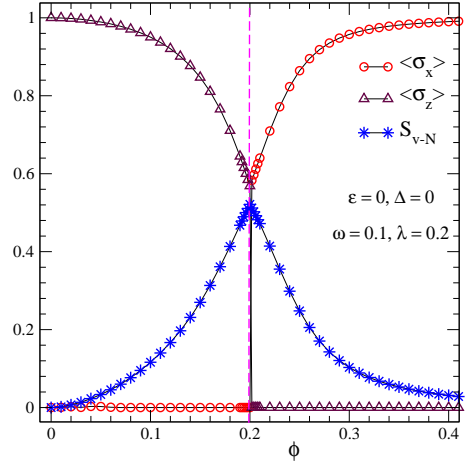}
\caption[FIG]{\label{SingMode00} (Color online) The magnetization $\langle{\sigma_z}\rangle$, spin coherence $\langle{\sigma_x}\rangle$ and entanglement entropy $S_{\rm v-N}$ are plotted as a function of the off-diagonal coupling $\phi$ for $\varepsilon=0,\Delta=0,\omega=0.1$ and $\lambda=0.2$. The dashed line marks sharp jumps of $\langle\sigma_z\rangle$ and $\langle\sigma_x\rangle$ at $\phi=0.2$. A truncated number of $N_{\rm tr}=40$ is used in Eqs. ~(\ref{Expand1}) and (\ref{Expand2}) in the exact diagonalization procedure. }
\end{figure}

Fig.~\ref{SingMode00} shows $\langle{\sigma_z}\rangle$ and $\langle{\sigma_x}\rangle$ as a function of the off-diagonal coupling strength $\phi$ for the diagonal coupling strength $\lambda=0.2$ and $\varepsilon=\Delta=0$. In this case, the ground state is always doubly degenerate as discussed before. With an increase (decrease) in the off-diagonal coupling strength $\phi$, $\langle\sigma_z\rangle$ ($\langle\sigma_x\rangle$) decays gradually until $\phi=\lambda=0.2$ where a sharp jump to zero occurs. $\langle\sigma_z\rangle$ ($\langle\sigma_x\rangle$) remains zero when $\phi$ is larger (smaller) than $\lambda$. However, the entanglement entropy $S_{\rm v-N}$ exhibits continuous behavior at $\phi=0.2$,
different from that of $\langle\sigma_x\rangle$ and $\langle\sigma_z\rangle$.

\subsection{Variational results}

For comparison, the two-bath model coupled to single-mode baths is also investigated by the variational approach.
A systematic coherent-state expansion of the ground state wave function is introduced as our variational ansatz \cite{Bera},
\begin{eqnarray}
|\Psi \rangle & = & |+\rangle \sum_{n=1}^{N} A_n \exp\left[ \sum_{l}^{M}\left(f_{n,l}b_l^{\dag} - \mbox{H}.\mbox{c}.\right)\right] |0\rangle_{\textrm{ph}} \nonumber \\
              & + & |-\rangle \sum_{n=1}^{N} B_n \exp\left[ \sum_{l}^{M}\left(g_{n,l}b_l^{\dag} - \mbox{H}.\mbox{c}.\right)\right] |0\rangle_{\textrm{ph}},
\label{vmwave}
\end{eqnarray}
where H$.$c$.$ denotes Hermitian conjugate, $|+\rangle$ ($|-\rangle$) stands for the spin up (down) state, and $|0\rangle_{\textrm{ph}}$
is the vacuum state of the boson bath. This ansatz describes a superposition of the localized states $|\pm\rangle$ which are correlated
to the effective bath modes with displacements $f_{n,l}$ and $g_{n,l}$, where $n$ stands for the $n$-th coherent state and $l$ denotes the $l$-th effective bath mode.
Since this trial wave function is identical to the Davydov ${\rm D}_1$ variational ansatz when $N=1$, it can also be termed as the
``multi-${\rm D}_1$ ansatz.'' In the single-mode case, the number of the effective bath modes $M$ is set to $1$.

The multi-${\rm D}_1$ ansatz is a generalization of the variational wave function originally proposed by Silbey and Harris \cite{Sil},
where the variational parameters are fixed to obey $A_n=B_n, f_{n,l}= -g_{n,l}$ and
$N=1$. It is also an extension of the hierarchy of translation-invariant ans{\"a}tze proposed by Zhao $et~al.$ \cite{zhao97}.
More than one coherent superposition states are considered in the multi-${\rm D}_1$ ansatz to capture bath entanglement,
and quantum fluctuations which are important to the quantum phase transition are well taken into account.
Theoretically, the number of coherent superposition states $N \rightarrow \infty$ is required for the completeness of the environmental wave function.
However, large values of $N$, pose significant challenges in carrying out numerical simulations.
We have found $N=4$ to be sufficient in obtaining reliable results for the variational approach, as the results from simulations with $N=6$ show no appreciable difference.

Using the multi-${\rm D}_1$ ansatz defined in Eq.~(\ref{vmwave}), the system energy $E$ can be calculated with
the Hamiltonian expectation $H=\langle\Psi|\hat{H}|\Psi\rangle$ and the norm of the wave function
$D=\langle\Psi|\Psi\rangle$ as $E=H/D$. The ground state is then obtained by minimizing
the energy with respect to the variational parameters $A_n, B_n, f_{n,l}$ and $g_{n,l}$.
The variational procedure entails $N(4M+2)$ self-consistency equations,
\begin{equation}
\frac{\partial H}{\partial x_{i}} - E\frac{\partial D}{\partial x_{i}} = 0,
\label{vmit}
\end{equation}
where $x_i(i=1,2,\cdots, 4NM+2N)$ denotes the variational parameters. For each set of the coefficients ($\alpha, \beta, s$ and $\bar{s}$)
of the continuous spectral densities defined in Eq.~(\ref{OspectraZ}), more than $100$ initial states are used in the iteration procedure with different sets of
variational parameters ($A_n, B_n$) uniformly distributed within an interval $[-1, 1]$.
The initial values of the parameters $f_{n,l}$ and $g_{n,l}$ are based on the classical displacements
to a minimum of the static spin-dependent potential, i.e., $f_{n,l} = -g_{n,l} \sim \lambda_{l} / 2\omega_{l}$
for the diagonal coupling bath, and $f_{n,l} = -g_{n,l} \sim \phi_l/2\omega_{l}$ for the off-diagonal coupling bath.
For the single-mode case, both of $f_{n,l}$ and $g_{n,l}$ as well as $A_n, B_n$ are initialized randomly.
After preparing the initial state, the relaxation iteration technique \cite{zhao92,zhao95} is adopted, and
simulated annealing algorithm is also employed to improve the energy minimization procedure. The iterative procedure is carried out until the target precision of $1\times10^{-14}$ is reached. Finer details of the variational approach are provided in Appendix~B.

With the ground state wave function $|\Psi_{\rm g} \rangle$ obtained so far, one can calculate the magnetization $\langle\sigma_z\rangle = \langle\Psi_{\rm g}|\sigma_z|\Psi_{\rm g}\rangle/D$, the spin coherence $\langle\sigma_x\rangle = \langle\Psi_{\rm g}|\sigma_x|\Psi_{\rm g}\rangle / D$ and the
ground state energy $E_{\rm g} = \langle\Psi_{\rm g}|\hat{H}|\Psi_{\rm g}\rangle/D$.
The von Neumann entropy $S_{\rm v-N}$ is also evaluated according to Eq.~(\ref{Entropy}). To further investigate the quantum phase transition, we introduce
the ground state fidelity $F$ \cite{zyy},
\begin{equation}
F(\beta) = \left|\langle \Psi_{\rm g}(\beta) | \Psi_{\rm g}(\beta')\rangle \right|/\sqrt{D(\beta)D(\beta')},
\label{vmfd}
\end{equation}
where $\beta'=\beta + \delta \beta$ is the neighboring Hamiltonian parameter, and $\delta \beta = 1\times 10^{-5}$.
An abrupt decrease in fidelity is expected to give a hint to the location of the transition, and a vanishing value of the fidelity
at the critical point $\beta_{\rm c}$ usually indicates a first-order phase transition.

\begin{figure}[tbp]
  \centering
  \includegraphics[width=0.8\linewidth]{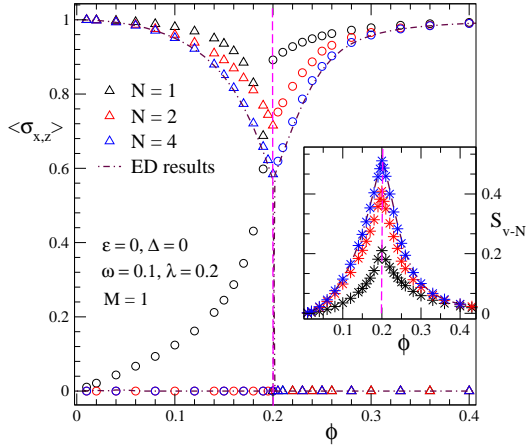}\\
  \caption{(Color online) The magnetization $\langle\sigma_{z}\rangle$ (triangles) and spin coherence $\langle\sigma_{x}\rangle$ (cirlces)
of two-bath model coupled to single-mode baths are displayed for various values of off-diagonal coupling strength $\phi$ in the case
of $\varepsilon=0,\Delta=0,\omega=0.1$ and $\lambda=0.2$.  In the inset, the von Neumann entropy $S_{\rm v-N}$ is plotted.
The symbols with black, red and blue colors, correspond to variational ans{\"a}tze with $N=1, 2$ and $4$,
respectively. The dashed lines mark the transition point, and the dash-dotted lines represent exact diagonalization results.}
\label{single_1}
\vspace{2\baselineskip}
\end{figure}

In Fig.~\ref{single_1}, the behavior of magnetization $\langle\sigma_{z}\rangle$, spin coherence $\langle\sigma_{x}\rangle$
and von Neumann entropy $S_{\rm v-N}$ is displayed for various values of off-diagonal coupling strength at
$\varepsilon=0,\Delta=0,\omega=0.1$ and $\lambda=0.2$. For comparison, exact diagonalization results for the same case
are also plotted with the dash-dotted lines. As $N$ increases,
the difference between variational and exact diagonalization results vanishes. It indicates that $N=4$ is sufficient to
reproduce the ground state of the two-bath model coupled to single-mode baths.

\begin{figure}[tbp]
  \centering
  \includegraphics[width=0.75\linewidth]{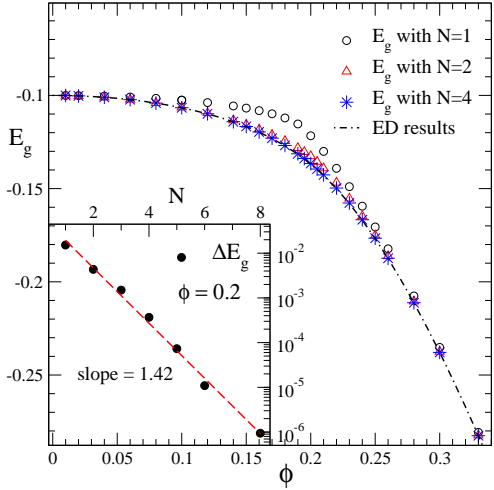}\\
  \caption{(Color online) The ground state energy $E_{\rm g}$ are shown for different
variational ans{\"a}tze with $N=1,2$ and $4$, in comparison with the exact diagonalization results. In the inset, the shift $\Delta E_{\rm g} = E_{\rm g}(N) - E_{\rm ed}$ at $\phi=0.2$
is displayed on a linear-log scale.  The dashed line represents an exponential fit. }
\label{single_2}
%\vspace{2\baselineskip}
\end{figure}

As shown in Fig.~\ref{single_2}, the ground state energy $E_{\rm g}$, equivalent to the free energy of the system, is also displayed for various numbers of coherence states $N$,
in comparison with exact diagonalization results. When $N=1$, i.e. the usual $\rm{D}_1$ ansatz, visible difference between variational and exact diagonalization results
can be observed nearby the transition point $\phi = 0.2$. It suggests that ${\rm D}_1$ ansatz is too simple to study the phase transition of the two-bath model, even with single-mode bosonic baths. The shift of the ground state energy $\Delta E_{\rm g}= E_{\rm g}(N) - E_{\rm ed}$ presented in the inset of Fig.~\ref{single_2} shows an exponential decay with $N$, and the slope of the linear-log plot ($1.42(6)$) is significantly large. We thus establish that, a small value of $N$, i.e., $N=4$, is sufficient to study the two-bath model via the variational approach.

\begin{figure}[htbp]
  \centering
  \includegraphics[width=0.85\linewidth]{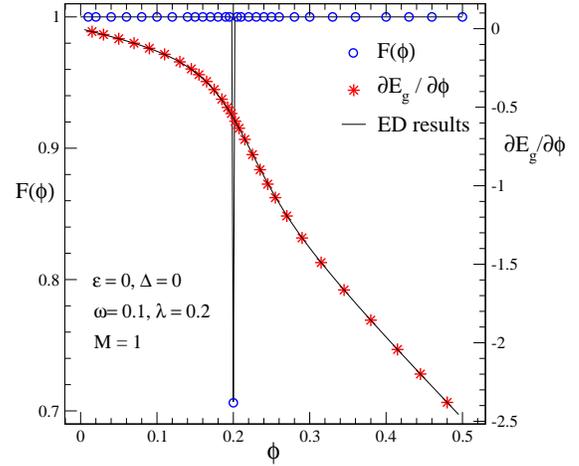}\\
  \caption{(Color online) The ground state fidelity $F$ and the derivative of the ground state energy $E_{\rm g}$ are displayed
 with respect to the off-diagonal coupling strength $\phi$ in the single-mode case with $\varepsilon=0,\Delta=0,\omega=0.1$ and $\lambda=0.2$.
 The open circles and stars represent variational results, and the solid lines denote the exact diagonalization results. }
\label{single_3}
\end{figure}

In Fig.~\ref{single_3}, the ground state fidelity $F(\phi)$ is plotted as a function of the off-diagonal coupling strength $\phi$.
A sharp decrease in $F(\beta)$ is observed at $\phi_{\rm c} = 0.2$, consistent with the exact diagonalization results.
The value of the fidelity $F(\phi_{\rm c})=0.706$ is much larger than zero, indicating that the transition is not of first order.
To confirm this contention, the derivative of the ground state energy $\partial E_{\rm g}/ \partial \phi$ is also displayed.
No discontinuity in $\partial E_{\rm g}/\partial \phi$ supports that the transition is softened, though the magnetization $\langle \sigma_z \rangle$ and spin coherence $\langle \sigma_x \rangle$ exhibit sharp jumps.

\section{Continuous spectral densities}

\subsection{Variational results}

\begin{figure}[tbp]
  \centering
  \includegraphics[width=0.85\linewidth]{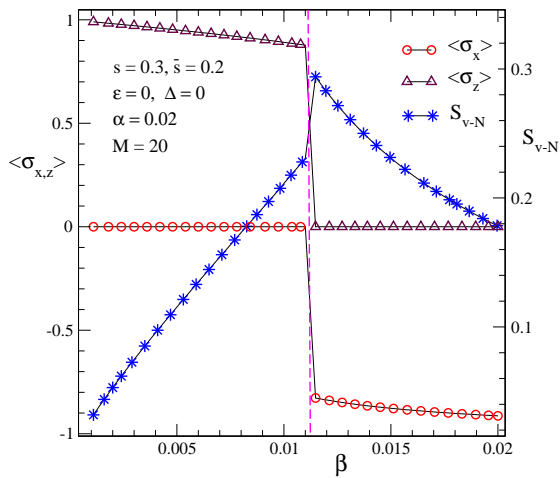}\\
  \caption{(Color online) The magnetization $\langle\sigma_{z}\rangle$, spin coherence $\langle\sigma_{x}\rangle$ and von Neumann
entropy $S_{\rm v-N}$ are plotted with respect to $\beta$ at $s=0.3$, $\bar{s}=0.2$ and $\alpha = 0.02$.
The dashed line indicates abrupt jumps. In both of the diagonal and off-diagonal coupling baths, $M=20$ is set. }
\vspace{2\baselineskip}
\label{vm1}
\end{figure}

The ground state properties of two-bath model with the baths described by a continuous spectral density function $J(\omega)$
are also studied via the variational approach. By adopting the logarithmic discretization procedure,
more than one effective bath modes are introduced in the variational calculations.
Fig.~\ref{vm1} shows the magnetization $\langle\sigma_z\rangle$, spin coherence $\langle\sigma_{x}\rangle$ and entanglement
entropy $S_{\rm v-N}$ plotted against the off-diagonal coupling strength ($\beta=0$ to $0.02$) when the number of effective bath modes $M=20$.  In these calculations, the other coefficients are set to $s=0.3, \bar{s}=0.2$ and $\alpha=0.02$. For simplicity, only one branch of the two-fold degenerate ground states is presented,
and the other can be obtained easily by projecting the operator ${\cal P}_{x}$ or ${\cal P}_{z}$ onto the ground state. Abrupt jumps are observed at $\beta \approx 0.011$ for all the three quantities, and such discontinuous behavior points to a first-order phase transition.
Since the ansatz employed in this work is much more sophisticated and contains more flexible variational parameters than the Silbey-Harris ansatz, it is important to distinguish the discontinuous behavior
shown in Fig.~\ref{vm1} from that obtained by Silbey-Harris variation in the biased single-bath model. The latter is regarded as an artifact arising from the excessive simplicity of the Silbey-Harris ansatz, which is considered poorly equipped to deal with the asymmetry induced by the bias \cite{Naz}.

\begin{figure}[tbp]
  \centering
  \includegraphics[width=0.85\linewidth]{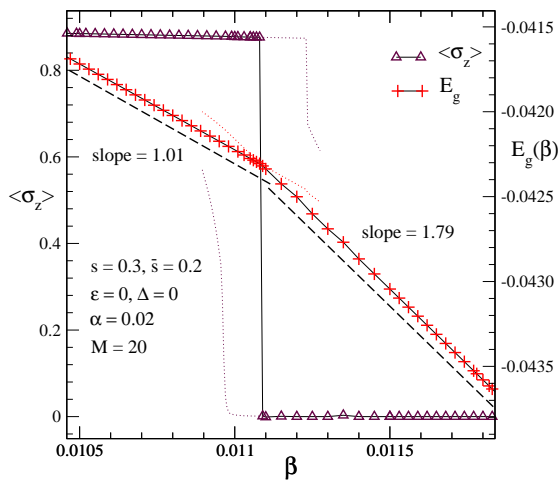}\\
  \caption{(Color online) The magnetization $\langle\sigma_{z}\rangle$ and ground state energy $E_{\rm g}$ are displayed in a small range of $\beta$
  at $s=0.3$, $\bar{s}=0.2$ and $\alpha = 0.02$. The critical point $\beta = 0.01109$ is located according to
  the discontinuity in the functions $\langle\sigma_{z}\rangle$ and $\partial E/\partial \beta$. The dotted lines represent the results
  of the metastable states. }
\label{vm2}
%\vspace{2\baselineskip}
\end{figure}

To locate the critical point more accurately, the magnetization $\langle \sigma_z \rangle$ is displayed in Fig.~\ref{vm2}
in a smaller range of $\beta$ from $0.0105$ to $0.012$. The transition point $\beta_{\rm c}=0.01109$ is then determined
according to the discontinuous behavior of $\langle \sigma_z \rangle$ within the interval $[0.01105, 0.01115]$.
Fig.~\ref{vm2} also shows the calculated ground state energy $E_{\rm g}$ as a function of $\beta$. Two different slopes resulting from linear fitting, $1.01$ and $1.79$, indicate that the derivative of the free energy $\partial E_{\rm g}/\partial \beta$ is discontinuous at the transition point, different from that in
the single-mode case shown in Fig.~\ref{single_3}. For comparison, the results for the metastable states are obtained from the relaxation iterations with gradually increasing (decreasing) off-diagonal coupling strength $\beta$, starting from the ground state at $\beta < \beta_{\rm c}$ ($\beta > \beta_{\rm c}$). After $\beta$ crosses the transition point, the system will be trapped in metastable states with higher system energy. It further supports the first-order nature of the phase transition.

\begin{figure}[tbp]
\vspace{0.7cm}
  \centering
  \includegraphics[width=0.9\linewidth]{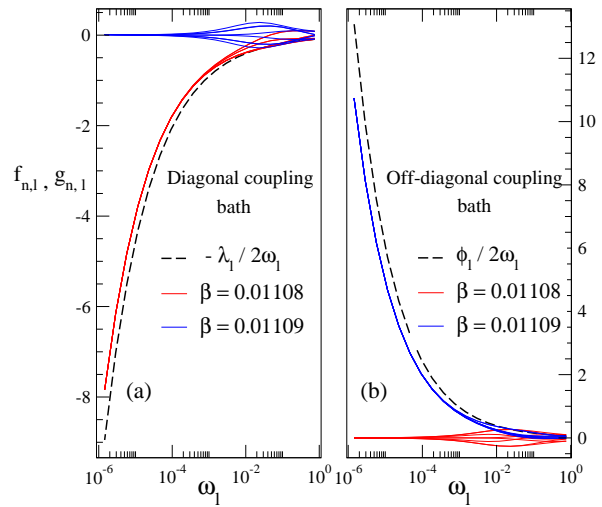} \\
  \caption{(Color online) The displacement  coefficients $f_{n,l}$ and $g_{n,l}$ determined at $\beta=0.01108$ and $0.01109$ are plotted
in (a) and (b) for the diagonal coupling and off-diagnoal coupling baths, respectively. The dashed lines represent
the classical displacements to the minimum of the static spin-dependent potential. }
\label{vm3}
\end{figure}

We next probe the wave function of the ground state in the vicinity of the transition point.
Shown in Fig.~\ref{vm3}(a) are the displacement coefficients $f_{n,l}$ and $g_{n,l}$ of the bath diagonally coupled to the spin for $\beta=0.01108$ and $0.01109$. For convenience, the notations $\Psi_{\rm A}$ and $\Psi_{\rm B}$ are used to denote wave functions of these two ground states.
At low frequencies, all the displacement coefficients converge to a value independent of $n$, i.e., $f_{n,l}=g_{n,l}\rightarrow -\lambda_{l} / 2\omega_{l}$ ($0$)
in $\Psi_{\rm A}$ ($\Psi_{\rm B}$). A huge jump appears in the low-frequency asymptotic value of the displacement coefficients as the coupling strength $\beta$ is changed by only a paltry amount of $10^{-5}$. A similar phenomenon can also be found in Fig.~\ref{vm3}(b)
for the displacements of the off-diagonal coupling bath, i.e., $f_{n,l}=g_{n,l} \approx 0$ in $\Psi_{\rm A}$
and $f_{n,l}= g_{n,l}\approx \phi_{l} / 2\omega_{l}$ in $\Psi_{\rm B}$.
At high frequencies, however, $f_{n,l}$ and $g_{n,l}$ exhibit quite different behavior not only at $\beta=0.01108$,
but also at $\beta=0.01109$ in both Fig.~\ref{vm3}(a) and (b).

\begin{figure}[tbp]
  \centering
  \includegraphics[width=0.85\linewidth]{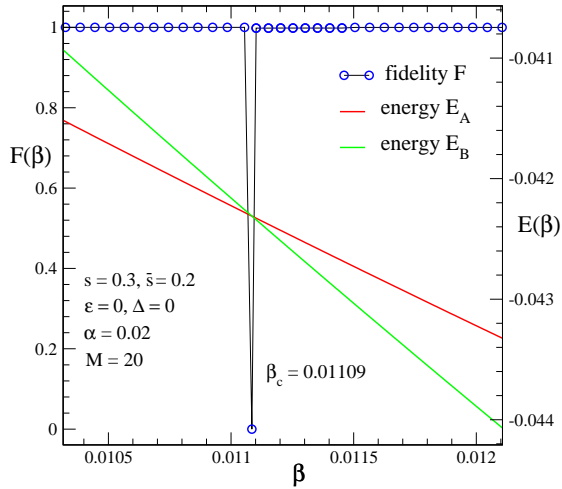} \\
  \caption{(Color online) The ground state fidelity $F$ and two energy functions $E_{\rm A}=\langle \Psi_{A}|\hat{H}|\Psi_{A}\rangle$
and $E_{\rm B}=\langle \Psi_{B}|\hat{H}|\Psi_{B}\rangle$ are plotted with $\beta$
in the case of $s=0.3$, $\bar{s}=0.2$ and $\alpha = 0.02$. At the critical point $\beta_{\rm c}$,
a sharp drop of $F(\beta)$ and an intersection of the two energy curves are presented.}
\label{vm4}
%\vspace{2\baselineskip}
\end{figure}

Fig.~\ref{vm4} shows the ground state fidelity $F(\beta)$ in the case of $s=0.3$, $\bar{s}=0.2$ and $\alpha = 0.02$.
A sharp drop in $F(\beta)$ at the critical point $\beta_{\rm c}=0.01109$ separates the ``localized
phase'' at small $\beta$ and ``delocalized phase'' at large $\beta$.  The vanishing value of the fidelity at $\beta=\beta_{\rm c}$, i.e., $F(\beta_{\rm c})=0$,
leads further support to the first-order transition. Since the fidelity maintains a value of unity on both sides of the transition point,
$\Psi_{\rm A}$ and $\Psi_{\rm B}$ shown in Figs.~\ref{vm3}(a) and (b) can be approximately considered as the ground states for $\beta < \beta_{\rm c}$
and $\beta > \beta_{\rm c}$, respectively. We further calculate the energies of the ground state and first excited state $E_{\rm A}=\langle \Psi_{A}|\hat{H}|\Psi_{A}\rangle$ and $E_B=\langle \Psi_{B}|\hat{H}|\Psi_{B}\rangle$. The fact that $E_{\rm A}$ and $E_{\rm B}$ exhibit a crossover at the critical point, is
consistent with the picture of the first-order phase transition, e.g., the ice-water phase transition \cite{Lan}.

\begin{figure}[tbp]
  \centering
  \includegraphics[width=0.85\linewidth]{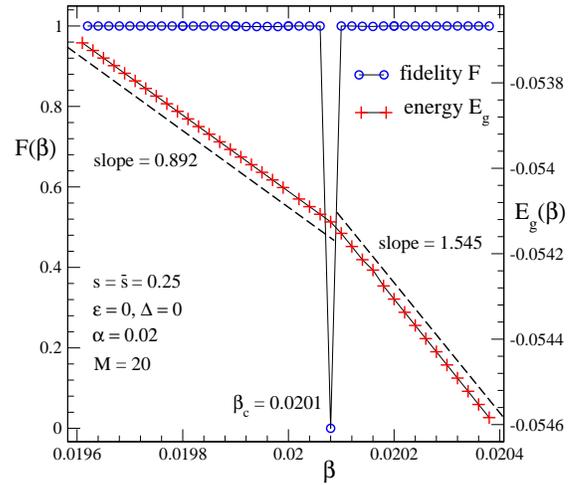} \\
  \caption{ (Color online) The ground state energy $E_{\rm g}$ and fidelity $F$ are displayed as a function of $\beta$
in the case of $s=\bar{s}=0.25$ and $\alpha = 0.02$. The discontinuity of the derivative of the energy curve $\partial E_{\rm g} /\partial \beta$
and a abrupt drop of the fidelity $F(\beta)$ are presented at $\beta_{\rm c}=0.0201$.}
\label{vm5}
\end{figure}

Finally, the case with two identical spectral exponents $s=\bar{s}$ is studied by the variational approach to further explore the competitive effects of the two phonon baths.
The ground state energy $E_{\rm g}(\beta)$ and fidelity $F(\beta)$ are displayed in Fig.~\ref{vm5},
for $s=\bar{s}=0.25$ and $\alpha = 0.02$.  According to the discontinuity in the derivative of the ground state energy $\partial E_{\rm g} /\partial \beta$
and the abrupt drop in the fidelity $F(\beta)$, one can locate the transition point $\beta_{\rm c}$ accurately. The resulting value of $\beta_{\rm c}=0.0201$ is in good agreement
with $\beta_{\rm c} = \alpha = 0.02$ obtained from the symmetry analysis. The relative error in the transition point is only $\delta \beta/\beta_{\rm c} = 0.5\%$. It thus indicates that the variational approach is an effective and feasible approach to study the quantum phase transition of the two-bath model.

\subsection{DMRG results}

To provide a platform for comparison of the results obtained by the variational approach, we have carried out DMRG calculations to investigate the quantum phase transition of the two-bath model.
Starting from Eq.~(\ref{sbm1_ctnu_h}) for the usual spin-boson model, one can map the phonon bath onto a Wilson chain by using the canonical transformation \cite{Wilson,chinmap}.
The Hamiltonian can be simultaneously mapped onto
\begin{eqnarray}
\hat{H}&=&\frac{\varepsilon}{2}\sigma_z-\frac{\Delta}{2}\sigma_x\nonumber\\
&+&\sum_{n=0} [\omega_n b_n^\dag b_n + t_n(b_n^\dag b_{n+1}+b_{n+1}^\dag b_n)]\nonumber\\
\label{wil_sbm1}
&+&\frac{\sigma_z}{2}\sqrt{\frac{\eta}{\pi}}(b^\dag_0+b_0).
\end{eqnarray}
where $b_n^\dag$ ($b_n$) are phonon creation (annihilation) operator, $\omega_n$ is the on site energy of site $n$, and $t_n$ is the hopping amplitude.
The coupling constant $\eta$ is proportional to $\alpha$, which is often chosen as the control parameter in the studies of the quantum phase transition of the spin-boson model.
Following the same routine of the single-bath spin-boson model, the two phonon baths in the two-bath model can be mapped onto two Wilson chains. The matrix product state (MPS) approach
is then adopted with an optimized phonon basis in the framework of DMRG to study the quantum phase transition in the ground state of the two-bath model.
The reader is referred to Appendix~C for detailed derivation of Hamiltonian mapping and introduction of MPS method.

\begin{figure}[tbp]
  \centering
  \includegraphics[width=1.0\linewidth]{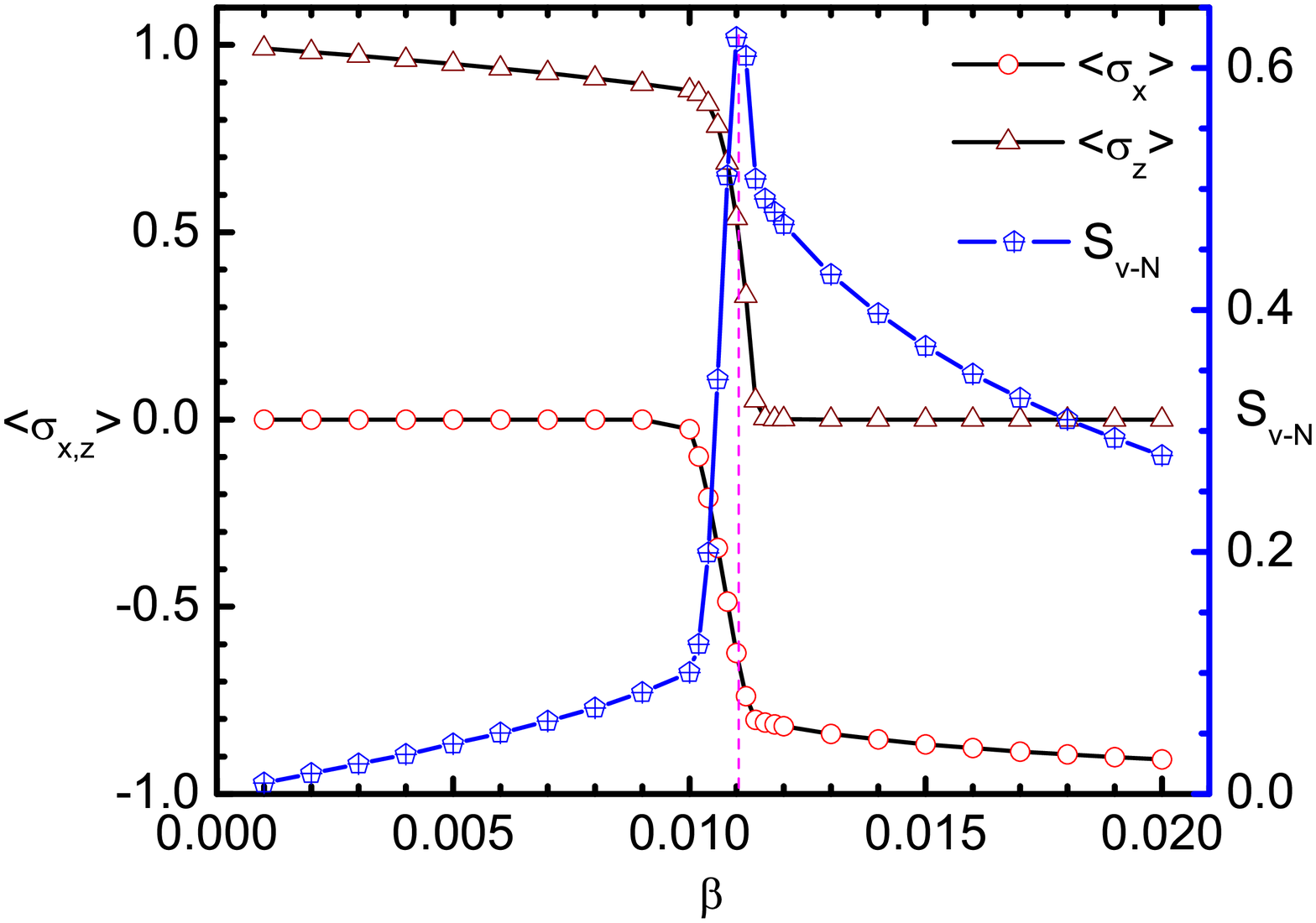}\\
  \caption{(Color online) $\langle\sigma_{x}\rangle$ and $\langle\sigma_{z}\rangle$ are plotted
at $s=0.3, \bar{s}=0.2$ and $\alpha=0.02$ with respect to $\beta$. The transition point is marked by the dashed line.
The corresponding von Neumann entropy $S_{\rm v-N}$ is shown as well with a sharp peak at the transition point $\beta_{\rm c}=0.0110$.}
\label{sns1}
\end{figure}

\begin{figure}[tbp]
  \centering
  \includegraphics[width=1.0\linewidth]{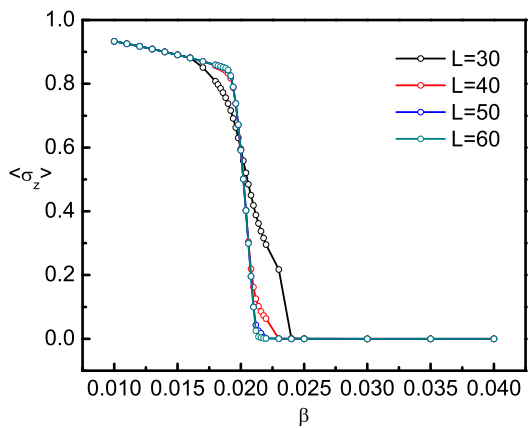}\\
  \caption{(Color online) The magnetization $\langle\sigma_{z}\rangle$ is displayed
at $s=\bar{s}=0.25$ and $\alpha=0.02$ with respect to $\beta$
for different lengths of Wilson chains $L=30, 40, 50$ and $60$.}
\label{sns2}
\end{figure}

For the convenience of comparison, the parameters $s=0.3, \bar{s}=0.2$ and $\alpha = 0.02$ are used in the DMRG calculations with $L=60, d_p=60, \Lambda=2$ and $D_c=50$
defined in Appendix~C. Fig.~\ref{sns1} shows $\langle\sigma_x\rangle, \langle\sigma_z\rangle$ and $S_{\rm v-N}$ in a range of $\beta$ from $0$ to $0.02$.
The transition point $\beta_{\rm c}=0.0110$ is determined by the peak of the entanglement entropy $S_{\rm v-N}$, in perfect agreement with the value of $0.01109$ found via variational calculations.  However, continuous behavior of $\langle \sigma_z \rangle, \langle \sigma_z \rangle$ and $S_{\rm v-N}$
is observed near the transition point, different from that observed in variational results shown in Figs.~\ref{vm1}, \ref{vm2} and \ref{vm4}.

The convergence of the finite size effect is investigated carefully in the DMRG calculations for various lengths of the Wilson chains $L=30, 40, 50$ and $60$.
Fig.~\ref{sns2} shows the magnetization $\langle\sigma_{z}\rangle$ with respect to $\beta$ for the case of $s=\bar{s}=0.25$ and $\alpha=0.02$. The parameters employed here are identical to those used in Fig.~\ref{vm5}. With an increase in $L$, the jump in the magnetization $\langle\sigma_{z}\rangle$ becomes increasingly sharper. For $L=60$, the transition point $\beta_{\rm c} =0.0202$ is determined, consistent with $\beta_{\rm c}=0.0201$ obtained by the variational approach.

\section{Discussion and conclusions}

At first glance, numerical results from the DMRG approach seem to suggest that the phase transition is continuous, different from first-order nature of the transition suggested by the variational results, in which, discontinuities are observed not only in $\langle\sigma_z\rangle, \langle\sigma_x\rangle$ and $S_{\rm v-N}$, but also in $F(\beta)$ and $\partial E_{\rm g}/\partial \beta$. The difference poses a question on whether the discontinuities uncovered are caused by artifacts arising from the variational approach.
According to the arguments on the results obtained for the single-mode case shown in Figs.~\ref{single_1}, \ref{single_2} and \ref{single_3},
the multi-${\rm D}_1$ variational ansatz with $N=4$ is sufficiently sophisticated to reproduce exact diagonalization results.
Furthermore, the convergence of $N$ is also investigated for the two-bath model coupled to the baths described by a continuous spectral function $J(\omega)$
in the case of $s=0.3, \bar{s}=0.2, \alpha=0.02$ and $\beta=0.011$. Correspondingly,
the ground state energies $E_{\rm g}=0.042255$ and $0.042283$ are obtained for $N=5$ and $6$, very close to the value of $0.042224$ at $N=4$.
To further verify that $N=4$ is sufficient to obtain reliable results for the quantum phase transition of the two-bath model, the transition point at $N=6$ is calculated to be $\beta_{\rm c}=0.01111$ in the case of $s=0.3, \bar{s}=0.2, \alpha=0.02$ and $M=20$. It is thus in good agreement with $\beta_{\rm c}=0.01109$
measured at $N=4$ in Fig.~\ref{vm2}.

\begin{figure}[tbp]
  \centering
  \includegraphics[width=0.73\linewidth]{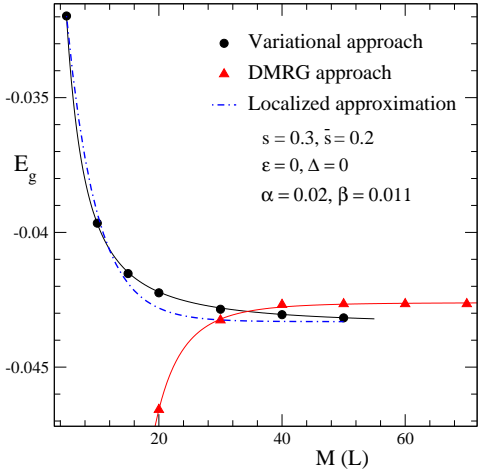} \\
  \caption{(Color online) The ground state energy $E_{\rm g}(M)$ is displayed with circles and triangles in the case
of $s=0.3, \bar{s}=0.2, \alpha=0.02$ and $\beta=0.011$. The solid lines represent power law fits $E_{\rm g}(M)= a M^{-b}+E_{\rm g}(\infty)$.
In the localized approximation, the energy of effective bath modes $\sum_l (-\lambda_l^2 / 4\omega_l)$ is
plotted with the dash-dotted line shifted up for the convenience of comparison.}
\label{vm6}
\end{figure}

The continuous behaviors of $\langle\sigma_z\rangle, \langle\sigma_x\rangle$ and $S_{\rm v-N}$
in Figs.~\ref{sns1} and \ref{sns2} may be misleading, since the numerical results of DMRG are sensitive to the boson number $d_p$, Wilson chain length $L$
and cut off dimension of the matrices $D_c$. As shown in Fig.~\ref{sns2}, the width of the transition regime
decays rapidly with an increase in the length $L$. Hence, it is reasonable to conjecture that the transition may be of first order in the limit of
$L,d_p,D_c \rightarrow \infty$. Moreover, similar linear behavior of the magnetization $\langle\sigma_z\rangle$ is observed apart from
the transition point in both of the DMRG and variational results shown in Figs.~\ref{vm1} and \ref{sns1}, consistent with the prediction of the
first-order transition theory \cite{bind}. Additional simulations with $s=\bar{s}=0.4, \alpha=0.1$ and $s=\bar{s}=0.6, \alpha=0.1$
are performed using the DMRG algorithm, and the discontinuity in the magnetization are found in both cases,
lending further support to the claim that the transition is of first order.

In addition, we have carefully examined the convergence of our results with respect to the effective bath-mode number $M$ for the variational approach. In Fig.~\ref{vm6}, the ground-state energy $E_{\rm g}$ is displayed as a function of $M$ in the case of $s=0.3, \bar{s}=0.2, \alpha=0.02$ and $\beta=0.011$. A power law decay curve of the form $E_{\rm g}(M)= a M^{-b}+E_{\rm g}(\infty)$ is found to provide a good fitting to the numerical data, which yields
the asymptotic value $E_{\rm g}(\infty)=-0.04345$. Since the length of the Wilson chain $L$ is equivalent to $M$,
the numerical results $E_{\rm g}(L)$ of DMRG for different values of $L$ are also shown in Fig.~\ref{vm6} for comparison.
The ground state energy from the variational approach is found to be lower than that from DMRG when $M$ and $L$ are sufficiently large, pointing to
the superiority of the variational results.

To further understand the decay of $E_{\rm g}(M)$, we focus on a localized bath state in which the energy of the $l$-th
effective bath mode can be approximated by $-\lambda_{l}^2/4\omega_{l}$ (derived from $f_{n,l},g_{n,l}=\pm \lambda_{l}/2\omega_{l}$).
According to the results presented in Fig.~\ref{vm3}, one bath of the two-bath model is in the localized state, and the other is in the delocalized state.
The contribution of the effective bath modes in the delocalized state to the ground state is neglectable since $f_{n,l},g_{n,l} \approx 0$.
Therefore, the energy of the effective bath modes can be calculated as $E_{\rm bath}=\sum_{l}^M(-\lambda_{l}^2/4\omega_{l})$.
As shown in Fig.~\ref{vm6}, $E_{\rm bath}(M)$ deceases with $M$ and tends to a constant value, in a trend similar to that of the ground state energy $E_{\rm g}(M)$. It indicates a change in the ground state of two-bath model by new effective bath modes, even though their frequencies are very low.

\begin{figure}[tbp]
  \centering
  \includegraphics[width=0.85\linewidth]{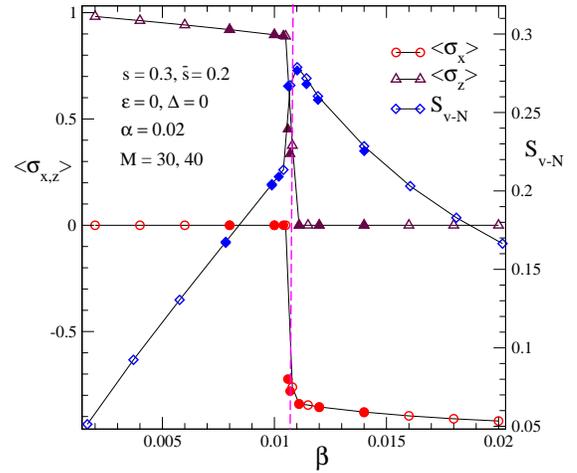} \\
  \caption{(Color online)  The magnetization $\langle\sigma_{z}\rangle$, spin coherence $\langle\sigma_{x}\rangle$ and entanglement entropy $S_{\rm v-N}$ are plotted
for $M=30$ (open symbols) and $40$ (solid symbols), respectively, in the case of
$s=0.3, \bar{s}=0.2$ and $\alpha=0.02$ with respect to $\beta$. The transition point is marked by the dashed line.}
\label{mod30}
\end{figure}

To investigate the influence of $M$ on the quantum phase transition of the two-bath model, we have carried out further simulations with $M=5,10,30$ and $40$ for
the case of $s=0.3, \bar{s}=0.2$ and $\alpha=0.02$ as an example. When $M\geq 10$, a first-order phase transition is observed,
and the transition point $\beta_{\rm c}(M)$ is determined accurately. For simplicity, only the results of $M=30$ and $M=40$ are displayed.
Fig.~\ref{mod30} shows the corresponding magnetization $\langle\sigma_{z}\rangle$, spin coherence $\langle\sigma_{x}\rangle$ and entanglement entropy $S_{\rm v-N}$. Sharp jumps in $\langle\sigma_{z}\rangle$, $\langle\sigma_{x}\rangle$ and $S_{\rm v-N}$
are observed, similar to the behavior shown in Fig.~\ref{vm1} at $M=20$.  However, the transition point $\beta_{\rm c} \approx 0.0108$
is much smaller than $\beta_{\rm c}=0.0111$ of $M=20$, thereby emphasizing the dependence of the critical point on $M$.

\begin{figure}[tbp]
  \centering
  \includegraphics[width=0.88\linewidth]{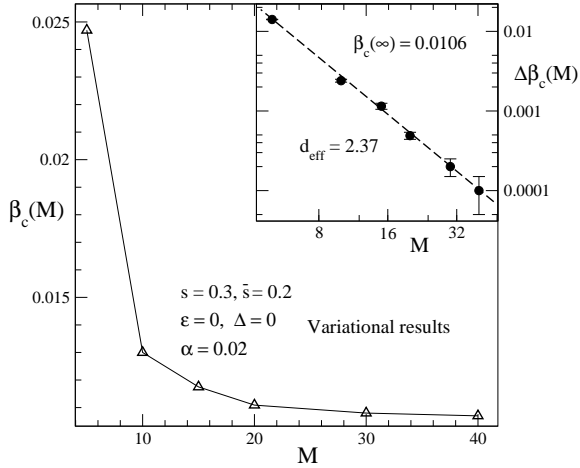} \\
  \caption{The transition point $\beta_{\rm c}$ determined by the variational results is displayed with open triangles as a function of the effective bath-mode number $M$.
Inset: the shift $\Delta \beta_{\rm c}(M)=\beta_{\rm c}(M) - \beta_{\rm c}(\infty)$ is plotted with solid circles on a log-log scale.
The dashed line represents a power law fit. }
\label{vm7}
\end{figure}

In order to reveal the relation $\beta_{\rm c}(M)$, the transition points calculated from the variational approach for different numbers of effective bath modes $M$ are depicted in Fig.~\ref{vm7}. With an increase in $M$, $\beta_{\rm c}$ is found to decrease monotonically tending to an asymptotic value $\beta_{\rm c}(\infty)$,
similar to the trend in $E_{\rm g}(M)$ shown in Fig.~\ref{vm6}. It suggests that the shift of the transition point $\beta_{\rm c}$ may possibly be related to the change in the ground state induced by the new effective bath modes.
General scaling arguments on the first order phase transition lead to a finite-size scaling relation \cite{Lan,Dic, Sin}
\begin{equation}
\Delta T_{\rm c}(L) = T_{\rm c}(L) - T_{\rm c}(\infty) \sim L^{-d},
\end{equation}
where $T_{\rm c}(\infty)$ is the transition point in the limit of $L \rightarrow \infty$, and $L^d$ is the system volume.
Similarly, $\Delta \beta_{\rm c}(M) = \beta_{\rm c}(M) - \beta_{\rm c}(\infty) \sim M^{-d_{\rm eff}}$ is assumed with an effective spatial dimension $d_{\rm eff}$
based on the equivalence between the number of effective bath modes $M$ and the length of Wilson chains $L$.
Taking $\beta_{\rm c}(\infty) = 0.0106$ as input, perfect power-law behavior of $\Delta \beta_{\rm c}(M)$ is presented in the inset of Fig.~\ref{vm7}.
From the slope, the effective spatial dimension is estimated as $d_{\rm eff}=2.37(6)$ for the two-bath model.
Interestingly, it is in good agreement with the prediction $d=3-2s=2.4$ by the mean-field analysis with $s=0.3$.

Finally, the symmetry analysis of two-bath model presented in section II is numerically verified. Taking the ground state $|\Psi_{\rm A}\rangle$ obtained
at $s=0.3, \bar{s}=0.02, \alpha=0.02, \beta=0.01108$ as an example, the influences of the symmetry operators ${\cal P}_{z},{\cal P}_{x}$
and ${\cal P}_{x}{\cal P}_{z}$ on the ground state are investigated, and the results are summarized in Tab.~\ref{t1}. Two-fold degenerate ground states
$|\Psi_{\rm A}\rangle$ (${\cal P}_{z}|\Psi_{\rm A}\rangle$) and ${\cal P}_{x}|\Psi_{\rm A}\rangle$ (${\cal P}_{x}{\cal P}_{z}|\Psi_{\rm A}\rangle$) are obtained according to
different values of the magnetization $\langle \sigma_z \rangle = \pm 0.87616$. The ground state energy $E_{\rm g}$, entropy $S_{\rm v-N}$ and spin coherence
$\langle \sigma_x \rangle$ of the two states are found to be nearly the same. The overlaps between $|\psi_{\rm A}\rangle,\ {\cal P}_{z}|\psi_{\rm A}\rangle,\ {\cal P}_{x}|\psi_{\rm A}\rangle,\
{\cal P}_{x}{\cal P}_{z}|\psi_{\rm A}\rangle,\ {\cal P}_{1}|\psi_{\rm A}\rangle$ and ${\cal P}_{2}|\psi_{\rm A}\rangle$ are also calculated.
The relations $\langle\psi_{\rm A}|{\cal P}_{x}{\cal P}_{z}{\cal P}_{x}|\psi_{\rm A}\rangle= -\langle\psi_{\rm A}|{\cal P}_{z}|\psi_{\rm A}\rangle= 1,\
\langle\psi_{\rm A}|{\cal P}_{z}{\cal P}_{2}|\psi_{\rm A}\rangle=\langle \sigma_z\rangle=-0.087617$ and
$\langle\psi_{\rm A}|{\cal P}_{2}|\psi_{\rm A}\rangle=|\langle \sigma_z\rangle|=0.087617$ are obtained along with $\langle\psi_{\rm A}|{\cal P}_{2}{\cal P}_{x}{\cal P}_{z}|\psi_{\rm A}\rangle = \langle\psi_{\rm A}|\sigma_{z}{\cal P}_{x}|\psi_{\rm A}\rangle = 0$ and $\langle\psi_{\rm A}|{\cal P}_{1}{\cal P}_{x}|\psi_{\rm A}\rangle= \langle\psi_{\rm A}|\sigma_{x}|\psi_{\rm A}\rangle=0$. All of them are consistent with the predictions of the symmetry analysis.
By projecting these operators onto another ground state $|\Psi_{\rm B}\rangle$ at the other side of the transition point,
similar properties are revealed except that the doubly degenerate ground states become $|\Psi_{\rm B}\rangle$ and ${\cal P}_{z}|\Psi_{\rm B}\rangle$ with different values of $\langle \sigma_x \rangle$. It further supports the contention that the phase transition does not remove the ground-state degeneracy, but rather eliminates spontaneous magnetization.

\begin{table}[ht]\centering
\caption{ The influences of the symmetry operations projecting onto the ground state $|\Psi_{\rm A}\rangle$ characterized by the magnetization $\langle\sigma_{z}\rangle$,
spin coherence $\langle\sigma_{x}\rangle$, von Neumann entropy $S_{\rm v-N}$ and ground state energy $E_{\rm g}$. }
\begin{tabular}[t]{l| c c c c}
\hline
\hline   States                           &    \quad $\langle \sigma_x\rangle$         &  $\langle \sigma_z\rangle$         &   $S_{\rm v-N}$         &    $E_{\rm g}$     \\
\hline
$|\Psi_{\rm A}\rangle$                          &    \quad $~8.4125$E$-11$   &  $-0.87616$  &  $0.23221$  &  $-4.2305$E$-2$       \\
${\cal P}_{z}|\Psi_{\rm A}\rangle$              &    \quad $-8.4125$E$-11$   &  $-0.87616$  &  $0.23221$  &  $-4.2305$E$-2$           \\
${\cal P}_{x}|\Psi_{\rm A}\rangle$              &    \quad $~8.4125$E$-11$   &  $~0.87616$  &  $0.23221$  &  $-4.2305$E$-2$        \\
${\cal P}_{x}{\cal P}_{z}|\Psi_{\rm A}\rangle$  &    \quad $-8.4125$E$-11$   &  $~0.87616$  &  $0.23221$  &  $-4.2305$E$-2$        \\
\hline \hline
\end{tabular}
\label{t1}
\end{table}

In summary, the ground state properties of the extended spin-boson model with two baths coupled to the $x$ and $z$ spin components, respectively, have been studied in this paper by the variational approach, the DMRG approach, exact diagonalization method as well as by the symmetry and mean-field analyses. A novel quantum phase transition from a doubly degenerate ``localized phase'' to the other doubly degenerate ``delocalized phase'' is uncovered.
Adopting the multi-${\rm D_1}$ ansatz as the variational wave function, transition points are determined accurately, in good agreement with the results of exact diagonalization and DMRG. According to the discontinuity in the magnetization, spin coherence, von Neumann entropy and  derivative of the ground state energy, and the vanishing value of the fidelity at the transition point, the transition is inferred to be of first order for the baths described by a continuous spectral density function.
In the case with single mode, however, the transition is found to be softened.
Furthermore, the convergence of results is carefully evaluated against different number of the coherence superposition states ($N$) and that of effective bath modes ($M$).
An effective spatial dimension is then calculated, consistent with the mean-field prediction within the error bar.

\section*{Acknowledgments}
The authors thank Yao Yao, Liwei Duan, Qinghu Chen and Bo Zheng for useful discussions.
Support from the Singapore National Research Foundation through the Competitive Research Programme (CRP) under Project No.~NRF-CRP5-2009-04
is gratefully acknowledged. The work is also supported in part by the National Natural Science Foundation of China under Grants No.~$11205043$.

\appendix
\section{The Ginzburg-Landau theory for the Spin-Boson model}
Consider a Spin-Boson (SB) Hamiltonian

\begin{equation}\label{sb}
\hat{H} = \Delta \sigma_x +\sigma_z\sum_k g_k x_k + \sum_k \left(\frac{p_k^2}{2m}+\frac{m\omega_k^2 x_k^2}{2}\right)  ,
\end{equation}
where coupling coefficients are characterized by the spectral function
\begin{equation}\label{spectrum}
J(\omega) = \sum_k g_k^2\delta(\omega_k - \omega) = \alpha \omega_c^{-s}\omega^{s}e^{-\frac{\omega}{\omega_c}}.
\end{equation}
We are primarily interested in the so-called subohmic regime with $0<s<1$. We also assume that the interaction cut-off frequency $\omega_c$ is sufficiently large, so that $\omega_c\gg \Delta$ and $\omega_c \gg \beta^{-1}$ ($\beta=T^{-1}$ is the inverse temperature).

We are interested in studying partition function of the system, $Z=Tr(e^{-\beta H})$. It is known that for a sufficiently large value of spin-bath coupling $\alpha$, the system exhibits a {\it continuous} phase transition into a localized phase, i.e., a phase where spin acquires spontaneous magnetization along the $z$-direction. (As a result, the oscillators shift from their equilibrium positions, e.g. some `weak' analogy with Pierles transition). Therefore one can introduce an order parameter as $B=\langle \sum_k g_k x_k \rangle$ by introducing a constraint in the Hamiltonian (\ref{sb}) through a Lagrange multiplier field $\lambda$. That is, we rewrite Eq. (\ref{sb}) as
\begin{equation}\label{heff}
H = \Delta \sigma_x +B \sigma_z + \sum_k \left(\frac{p_k^2}{2m}+\frac{m\omega_k^2 x_k^2}{2}\right) + i\lambda (B-\sum_k g_k x_k).
\end{equation}
By integrating out the oscillators and then the field $\lambda$, we obtain an (imaginary time) effective action for the spin+order parameter system,
\begin{equation}\label{seff}
S= \int_0^\beta d\tau \left(\Delta \sigma_x +B(\tau) \sigma_z\right) + \frac{1}{2\beta}\sum_{\omega_n}\frac{|B(\omega_n)|^2}{K(\omega_n)},
\end{equation}
where $\omega_n=2\pi n/\beta$ are Matsubara frequencies and
\begin{equation}\label{k}
K(\omega)=\sum_k \frac{g_k^2}{m(\omega^2+\omega_k^2)}=\int d\omega^\prime \frac{J(\omega^\prime)}{m(\omega^2+{\omega^\prime}^2)}.
\end{equation}
Note that for the spectral function given by Eq. (\ref{spectrum}), $K(\omega) \simeq K(0) - c\omega^s$ for $\omega\ll\omega_c$, where $c=(\frac{\alpha}{m \omega_c^s})\int_0^\infty x^{s-1}(1+x^2)^{-1}dx$, and  $K(\omega) \sim \omega^{-2}$ for $\omega\gg\omega_c$.

In order to average over the spin, one needs to evaluate the time-ordered exponent ${\cal T} e^{-\int_0^\beta d\tau \left(\Delta \sigma_x +B(\tau) \sigma_z\right)}$. This can be done perturbatively in $B$. Indeed, near the critical point the value of the order parameter $B$ is infinitesimally small, and therefore such expansion is well justified. Furthermore, near the phase transition point, the energy functional Eq.~(\ref{seff}) is dominated by low frequency fluctuations. Therefore for finite $\Delta$, when evaluating the time ordered exponent, one can use an adiabatic approximation (in $B(\tau)$). For sufficiently high $\beta$ (i.e. low temperature),
\begin{equation}\label{toexp}
Tr_{\rm spin}\left({\cal T} e^{{-\int_0^\beta }d\tau \left(\Delta \sigma_x +B(\tau) \sigma_z\right)}\right)\simeq e^{\int_0^\beta d\tau \sqrt{\Delta^2+B^2(\tau)}}  ,
\end{equation}
where we have dropped the term with $+\sqrt{\Delta^2+B^2(\tau)}$ eigenvalue. Expanding the square-root up to the quartic order in $B$, we obtain an effective Ginzubrg-Landau-type functional for the partition function of the system in the vicinity of critical point, $Z=\int {\cal D}B(\tau)\,e^{-F_{\rm eff}}$, where
\begin{equation}\label{seff0}
F_{\rm eff}= \frac{1}{2\beta}\sum_{\omega_n}(\frac{1}{K(\omega_n)}-\frac{1}{\Delta}) |B(\omega_n)|^2 +\frac{1}{4\Delta^3}
\int d\tau |B(\tau)|^4.
\end{equation}
Note that the use of the adiabatic approximation was not necessary. An explicit account of non-locality in the quadratic term gives
\begin{eqnarray}\label{seff1}
F_{\rm eff} & = & \frac{1}{2\beta}\sum_{\omega_n}(\frac{1}{K(\omega_n)}-\frac{\Delta}{\Delta^2+\omega_n^2}) |B(\omega_n)|^2  \nonumber \\
& + & \frac{1}{4\Delta^3} \int d\tau |B(\tau)|^4,
\end{eqnarray}
where, in the spirit of Ginzburg-Landau expansion, the frequency dependence in the quartic term is neglected.
The energy functional exhibits an instability given at the mean field level by the condition $\Delta=K(0)$. Note that  the phase transition is believed to occur at the critical value of $\alpha_{\rm c} \sim \Delta^{1-s}$, while our mean field condition corresponds to $\alpha_{\rm c} \sim \Delta$. So, presumably, the renormalization effects are strong for $s\sim 1$.

To the first order, the renormalization of the phase transition point (i.e., one loop correction) is given by the equation
\begin{equation}\label{oneloop}
\frac{1}{K(0)}-\frac{1}{\Delta} + \frac{1}{2\Delta^3}\int \frac{d\omega}{2\pi} \frac{1}{\frac{1}{K(\omega)}-\frac{\Delta}{\Delta^2+\omega^2}}=0 .
\end{equation}
Here, for simplicity, we consider $T=0$ case.

\section{The variational approach}

For convenience, the Hamiltonian in Eq.~(\ref{Ohami}) can be recast in a single-bath form
\begin{eqnarray}
\hat{H} & = &-\frac{\Delta}{2} \sigma_x + \frac{\varepsilon}{2}\sigma_z +\sum_{l}\omega'_l{b'_l}^{\dag}b'_l  \\ \nonumber
& + & \frac{\sigma_z}{2}\sum_{l}\lambda'_l\left( b'_l + {b'_l}^{\dag}\right) + \frac{\sigma_x}{2}\sum_{l}\phi'_l\left( b'_l + {b'_l}^{\dag}\right),
\label{vmhami}
\end{eqnarray}
by the transformation
\[
\omega'_l  = \left\{
   \begin{array}{lll}
    \omega_l,         & \quad & \mbox{ $0 < l \leq M $} \\
    \omega_{l-M},     & \quad & \mbox{ $M < l \leq 2M $}
   \end{array}, \right.
\]
\[
\lambda'_l = \left\{
   \begin{array}{lll}
    \lambda_l,        & \quad & \mbox{  $0 < l \leq M $} \\
    0,                & \quad & \mbox{  $M < l \leq 2M $}
   \end{array}, \right.
\]
\[
\phi'_l  = \left\{
   \begin{array}{lll}
    0,                & \quad & \mbox{ $0 < l \leq M $} \\
    \phi_{l-M},       & \quad & \mbox{ $M < l \leq 2M $}
   \end{array}, \right.
\]
\begin{equation}
b'_l  = \left\{
   \begin{array}{lll}
    b_{l, 1},         & \quad & \mbox{ $0 < l \leq M $} \\
    b_{l-M, 2},       & \quad & \mbox{ $M < l \leq 2M$}
   \end{array}, \right.
\label{vmtran}
\end{equation}
where $M$ is the number of effective bath modes for both of the diagonal and off-diagonal coupling baths.
Using the multi-${\rm D}_1$ ansatz defined in Eq.~(\ref{vmwave}) as trial wave function, the system energy can be calculated
as $E=H/D$, where $H$ is the Hamiltonian expectation and $D$ is the normal of the wave function. In the case $\varepsilon=\Delta=0$,
they can be written as
\begin{eqnarray} H&=& \sum_{m,n}A_mB_n\Gamma_{m,n}\sum_{k}\eta_k(f_{m,k}+g_{n,k})   \\
&+& \sum_{m,n}A_mA_n F_{m,n}\sum_{k}\left[\omega_{k}f_{m,k}f_{n,k} + \frac{\lambda_{k}}{2}(f_{m,k}+f_{n,k})\right] \nonumber \\
&+& \sum_{m,n}B_mB_n G_{m,n}\sum_{k}\left[\omega_{k}g_{m,k}g_{n,k} - \frac{\lambda_{k}}{2}(g_{m,k}+g_{n,k})\right], \nonumber
\label{vmH}
\end{eqnarray}
and
\begin{equation}
D=\langle \psi |\psi\rangle = \sum_{m,n}\left( A_mA_nF_{m,n} + B_mB_nG_{m,n}\right),
\label{vmD}
\end{equation}
where $F_{m,n},G_{m,n}$ and $\Gamma_{m,n}$ are Debye-Waller factors defined as
\begin{eqnarray}
F_{m,n} & = & \exp\left[-\frac{1}{2}\sum_{k}(f_{m,k}-f_{n,k})^2\right], \nonumber \\
G_{m,n} & = & \exp\left[-\frac{1}{2}\sum_{k}(g_{m,k}-g_{n,k})^2\right], \nonumber \\
\Gamma_{m,n} & = & \exp\left[-\frac{1}{2}\sum_{k}(f_{m,k}-g_{n,k})^2\right].
\label{vmfactor}
\end{eqnarray}

One can get a set of self-consistency equations with the form of Eq.~(\ref{vmit}) by minimizing the energy $E=H/D$ with respect to the variational parameters.
They can also be deduced by the Lagrange multiplier method when we consider the constraint condition $D\equiv1$. Finally, the iterative equations are derived
{\small
\begin{eqnarray}
A_n^{*} & = &  \frac{\sum_{m}B_m\Gamma_{n,m}dd_{n,m}+2\sum_m^{m\neq n}A_mF_{n,m}(aa_{n,m}-E)}{2(E-a_{n,n})}, \nonumber \\
B_n^{*} & = &  \frac{\sum_{m}A_m\Gamma_{m,n}dd_{m,n}+2\sum_m^{m\neq n}B_mG_{m,n}(bb_{m,n}-E)}{2(E-b_{n,n})},  \nonumber \\
f_{m,k}^{*} & = & \frac{2\sum_n^{n\neq m} A_nF_{m,n}(\omega_kf_{n,k}+\lambda_k/2+aa_{m,n}f_{n,k}-Ef_{n,k})}{2A_m(E-\omega_k-aa_{m,m})}  \nonumber \\
        & + &  \frac{\sum_nB_n\Gamma_{m,n}(g_{n,k}dd_{m,n} + \eta_{k})+ A_m\lambda_k}{2A_m(E-\omega_k-aa_{m,m})},  \nonumber \\
g_{m,k}^{*} & = & \frac{2\sum_n^{n\neq m} B_nG_{m,n}(\omega_kg_{n,k}-\lambda_k/2+bb_{m,n}g_{n,k}-Eg_{n,k})}{2B_m(E-\omega_k-bb_{m,m})}  \nonumber \\
        & + &  \frac{\sum_nA_n\Gamma_{n,m}(f_{n,k}dd_{n,m} + \eta_{k})- B_m\lambda_k}{2B_m(E-\omega_k-bb_{m,m})},
\label{vmit2}
\end{eqnarray}
}
where ${\rm dd}_{m,n},{\rm aa}_{m,n}$ and ${\rm bb}_{m,n}$ denote
\begin{eqnarray}
dd_{m,n} & = & \sum_k\eta_k(f_{m,k} + g_{n,k}), \nonumber \\
aa_{m,n} & = & \sum_k\left[\omega_kf_{m,k}f_{n,k} + \frac{\lambda_k}{2}(f_{m,k}+f_{n,k})\right],   \nonumber \\
bb_{m,n} & = & \sum_k\left[\omega_kg_{m,k}g_{n,k} - \frac{\lambda_k}{2}(g_{m,k}+g_{n,k})\right],
\label{vmfactor2}
\end{eqnarray}
respectively. Using the relaxation iteration technique, one updates the variation parameters by $x_i'=x_i+t*(x_i^{*}-x_i)$,
where $x_i^{*}$ is defined in Eq.~(\ref{vmit2}), and $t$ is the relaxation factor.
In usual, $t=0.1$ is set in the variational procedure, while it gradually decreases to $0.001$ in the
simulated annealing algorithm.
With the ground state at hand, the magnetization $\langle \sigma_z\rangle$ and spin coherence $\langle \sigma_x\rangle$
can be calculated by
\begin{eqnarray}
\langle \sigma_z\rangle & = & \frac{\sum_{m,n}A_mA_nF_{m,n}-B_mB_nG_{m,n}}{\sum_{m,n}A_mA_nF_{m,n}+B_mB_nG_{m,n}}, \nonumber \\
\langle \sigma_x\rangle & = & \frac{\sum_{m,n}2A_mB_n\Gamma_{m,n}}{\sum_{m,n}A_mA_nF_{m,n}+B_mB_nG_{m,n}}.
\label{vmpro}
\end{eqnarray}
And the entanglement entropy $S_{\rm v-N}$ and ground state energy $E_{\rm g}$ are measured
according to Eq.~(\ref{Entropy}) and $H/D$, respectively.

\section{The DMRG Method}

In order to deal with the two-bath model by employing the DMRG algorithm, followed by the standard theoretical treatment \cite{Wilson,Bulla,chinmap} that leads to
Eq.~(\ref{wil_sbm1}), the two phonon baths are transformed into two Wilson chains. The Hamiltonian Eq.~(\ref{Ohami}) is mapped simultaneously to
\begin{eqnarray}
\hat{H}&=&\frac{\varepsilon}{2}\sigma_z-\frac{\Delta}{2}\sigma_x\nonumber\\
&+&\sum_{n=0,i} [\omega_{n,i} b_{n,i}^\dag b_{n,i} + t_{n,i}(b_{n,i}^\dag b_{n+1,i}+b_{n+1,i}^\dag b_{n,i})]\nonumber\\
%&+&\sum_{n=0} [\omega_n b_n^\dag b_n + t_n(b_n^\dag b_{n+1}+b_{n+1}^\dag b_n)]\nonumber\\
\label{W_ham}
&+&\frac{\sigma_z}{2}\sqrt{\frac{\eta_z}{\pi}}(b^\dag_{0,1}+b_{0,1})+\frac{\sigma_x}{2}\sqrt{\frac{\eta_x}{\pi}}(b^\dag_{0,2}+b_{0,2}),
\end{eqnarray}
where
\begin{equation}\label{etax}
  \eta_{x} = \int_{0}^{\omega_c}J_{x}(\omega)d\omega=\frac{2\pi\beta}{1+\bar{s}}\omega^2_c,
\end{equation}
\begin{equation}\label{etaz}
  \eta_{z} = \int_{0}^{\omega_c}J_{z}(\omega)d\omega=\frac{2\pi\alpha}{1+s}\omega^2_c,
\end{equation}
and $i = 1, 2$ is the index for the baths. $\omega_{n,i}$ and $t_{n,i}$ are given as \cite{chinmap},
\begin{equation}\label{omega}
  \omega_{n,1} = \zeta_s(A_n+C_n),
\end{equation}
\begin{equation}\label{tn}
  t_{n,1} = -\zeta_s(\frac{N_{n+1}}{N_n})A_n,
\end{equation}
\begin{equation}\label{zetas}
  \zeta_s = \omega_c \frac{s+1}{s+2}\left(\frac{1-\Lambda^{-(s+2)}}{1-\Lambda^{-(s+1)}}\right),
\end{equation}
\begin{equation}\label{An}
  A_n = \Lambda^{-n}\frac{(1-\Lambda^{-(n+1+s)})^2}{(1-\Lambda^{-(2n+1+s)})(1-\Lambda^{-(2n+2+s)})},
\end{equation}
\begin{equation}\label{Cn}
  C_n = \Lambda^{-(n+s)}\frac{(1-\Lambda^{-n})^2}{(1-\Lambda^{-(2n+s)})(1-\Lambda^{-(2n+1+s)})},
\end{equation}
\begin{equation}\label{Nn}
  N^2_n = \Lambda^{-n(1+s)}\frac{(\Lambda^{-1};\Lambda^{-1})_{n}^2}{(\Lambda^{-(s+1)};\Lambda^{-1})_{n}^2(1-\Lambda^{-(2n+1+s)})},
\end{equation}
with
\begin{equation}\label{poly}
  (a;b)_n = (1-a)(1-ab)(1-ab^2)\cdots(1-ab^{(n-1)})
\end{equation}
and the discretization parameter $\Lambda=2$.

In the Fock representation, the ground state wave function of Hamiltonian~(\ref{W_ham}) characterizing a single chain system can be written in the form of matrix-product states (MPS) as
\begin{equation}\label{mps}
  |\psi\rangle = \sum_{i_0 = \uparrow,\downarrow; {j}}A^{i_0}A^{j_1}A^{j_2}\cdots A^{j_{L-1}}|i_0,\vec{j}\rangle,
\end{equation}
where $i_0$ is the spin index, $\vec{j} = (j_1,j_2,\cdots j_{L-1})$, $0\leq j_i\leq d_p$, represents the quantum numbers for the phonon basis, $L$ is the length of the chain,
and $d_p$ is the the number of phonon allocated on each site on the chain. ${A^j}$ defined in Eq.~(\ref{mps}) are single matrices whose dimension is restricted by a cut off $D_c$.

Subsequently, performing the iterative optimization procedure \cite{mps_c}, each matrix $A$ can be optimized with the truncation error less than $10^{-7}$. Furthermore, if we used the DMRG algorithm combined with the optimized phonon basis \cite{obb,guo}, the phonon numbers $d_p$ on each site of the Wilson chain can be kept up to $100$. Therefore, totally about $10^2L$ phonons will be included in the DMRG calculations. Here, in order to determine the phase transition conclusively, at least $d_p=60$ phonon should be kept in the calculation. After that, $\langle\sigma_{x}\rangle$, $\langle\sigma_{z}\rangle$ and the von-Neumann entropy
\begin{equation}\label{vN}
  S_{\textrm{v-N}} = -\textrm{Tr}\rho_s\textrm{log}\rho_s,
\end{equation}
where $\rho_s$ is the reduced density matrix of the spin can all be extracted by performing common quantum averaging using the MPS wavefunction.

%Unused bibitems

%\bibitem{ther} T. Ruokola, and T. Ojanen, Phys. Rev. B. \textbf{83}, 045417 (2011).
%\bibitem{qs4} R. J. Schoelkopf and S. M. Girvin, Nature (London), \textbf{451}, 664 (2008).
%\bibitem{amo} P. Esquinazi, \textsl{Tunneling Systems in Amorphous and Crystalline Solids}, (Springer-Verlag, Berlin, Heidelberg, 1998)
%\bibitem{Khveshchenko} D. V. Khveshchenko, Phys. Rev. B. {\bf 69}, 153311 (2004); M. Thorwart and P. Hanggi, Phys. Rev. A. {\bf 65}, 012309 (2002).
%\bibitem{dc2} I. V. Lerner, B. L. Altshuler, and Y. Gefen, eds., \textsl{Fundamental Problems of Mesoscopic Physics: Interactions and Decoherence}, Nato Science Series, Vol. 154 (Kluwer Academic Publishers, 2004).
%\bibitem{Marcus}R. A. Marcus and N. Sutin, Biochim. Biophys. Acta. \textbf{811},265 (1985).
%\bibitem{dc1} W. H. Zurek, Rev. Mod. Phys., \textbf{75}, 715 (2003).


\begin{thebibliography}{999}
\bibitem{Leggett} A. J. Leggett, S. Chakravarty, A. T. Dorsey, M. P. A. Fisher, A. Garg, and W. Zwerger, Rev. Mod. Phys. \textbf{59}, 1 (1987).





\bibitem{Weiss} U.~Weiss, {\it Quantum Dissipative Systems}, 3rd ed. (World Scientific, Singapore, 2007).





\bibitem{qs1} Y. Makhlin, G. Schon, and A. Shnirman, Rev. Mod. Phys., \textbf{73}, 357 (2001).





\bibitem{qs2} D. Vion, A. Aassime, A. Cottet, P. Joyez, H. Pothier, C. Urbina, D. Esteve, and M. H. Devoret, Science, \textbf{296}, 886 (2002).





\bibitem{qs3} J. Koch, T. M. Yu, J. Gambetta, A. A. Houck, D. I. Schuster, J. Majer, A. Blais, M. H. Devoret, S. M. Girvin, and R. J. Schoelkopf, Phys. Rev. A, \textbf{76}, 042319 (2007).





\bibitem{dyna} Y. Yao, L. Duan, Z. Lu, C. Q. Wu, and Y. Zhao, Phys. Rev. E. \textbf{88}, 023303 (2013).





\bibitem{dua}  L. Duan, H. Wang, Q. H. Chen, and Y. Zhao,  J. Chem. Phys. \textbf{139}, 044115 (2013).




\bibitem{et1} A. Garg, J. N. Onuchic, and V. Ambegaokar, J. Chem. Phys. \textbf{83}, 4491 (1985).





\bibitem{et2} L. M{\"u}hlbacher and R. Egger, J. Chem. Phys. \textbf{118}, 179 (2003); Chem. Phys. \textbf{296}, 193 (2004).





\bibitem{qpt1} S. K. Kehrein and A. Mielke, Phys. Lett. A. \textbf{219}, 313 (1996).





\bibitem{qpt2} M. Vojta, N. Tong, and R. Bulla, Phys. Rev. Lett. \textbf{94}, 070604 (2005).





\bibitem{alv} A. Alvermann and H. Fehske, Phys. Rev. Lett. \textbf{102}, 150601 (2009)





\bibitem{win}  A. Winter, H. Rieger, M. Vojta, and R. Bulla, Phys. Rev. Lett.  \textbf{102}, 030601 (2009).




\bibitem{Holstein} T. Holstein, Ann. Phys. \textbf{8}, 325 (1959); \textbf{8}, 343 (1959).





\bibitem{Su} W. P. Su, J. R. Schrieffer, and A. J. Heeger, Phys. Rev. Lett. \textbf{42}, 1698 (1979).





\bibitem{lv} Z. Lv, L. Duan, X. Li, P. M. Shenai, and Y. Zhao, J. Chem. Phys. \textbf{139}, 164103 (2013).





\bibitem{pach}  L. A. Pach\'{o}n and P. Brumer, Phys. Rev. A. \textbf{87}, 022106 (2013) 

\bibitem{you}   J. Q. You and F. Nori, Physics Today, \textbf{58}, 42 (2005)


\bibitem{card}  P. C. C\'{a}rdenas, M. Paternostro, and F. L. Semi\~{a}o, preprint arXiv:1406.4899 (2014)


\bibitem{raft}  J. Raftery, D. Sadri, S. Schmidt, H. E. T\"{u}reci, and A. A. Houck, Phys. Rev. X. \textbf{4}, 031043 (2014)


\bibitem{liao}  J. Q. Liao and L. M. Kuang, J. Phys. B: At. Mol. Opt. Phys. \textbf{40}, 1845 (2007)


\bibitem{ruok}  T. Ruokola and T. Ojanen, Phys. Rev. B. \textbf{83}, 045417 (2011)

\bibitem{guo} C. Guo, A. Weichselbaum, J. V. Delft, and M. Vojta, Phys. Rev. Lett.  \textbf{108}, 160401 (2012).


\bibitem{Bulla} R. Bulla, N.-H. Tong and M. Vojta, Phys. Rev. Lett. \textbf{ 91}, 170601 (2003); M. Vojta, N.-H. Tong, and R. Bulla, Phys. Rev. Lett.  \textbf{94}, 070604 (2005);
F. B. Anders, R. Bulla and M. Vojta, Phys. Rev. Lett.  \textbf{98}, 210402 (2007).





\bibitem{bulla2} R. Bulla, H. J. Lee, N. H. Tong, and M. Vojta, Phys. Rev. B  \textbf{71}, 045122 (2005).




\bibitem{Costi} T. A. Costi and R. H. Mckenzie. Phys. Rev. A.  \textbf{68},034301 (2003); K. L. Hur, P. D. Beaupre, and W. Hofstetter. Phys. Rev. Lett.  \textbf{99},126801 (2007).





\bibitem{zyy} Y. Y. Zhang, Q. H. Chen, and K. L. Wang, Phys. Rev. B.  \textbf{81},121105(R) (2010)





\bibitem{Wu}   N. Wu, L. Duan, X. Li, and Y. Zhao, J. Chem. Phys.  \textbf{138}, 084111 (2013).




\bibitem{Chen} Q. H. Chen, Y. Y. Zhang, T. Liu, and K. L. Wang, Phys. Rev. A. \textbf{78}, 051801(R) (2008)




\bibitem{chinmap} A. W. Chin, A. Rivas, S. F. Huelga, and M. B. Plenio, J. Math. Phys. \textbf{51}, 092109 (2010).





\bibitem{Bennett} C. H. Bennett, D. P. Divincenzo, J. A. Smolin, W. K. Wootters. Phys. Rev. A.  \textbf{54}, 3824 (1996); M. A. Nielsen and I. L. Chuang, \textit{Quantum Computation and Quantum Information} (Cambridge University Press, Cambridge, England, 2004).





\bibitem{Zhao} Y. Zhao, P. Zanardi, and G. H. Chen. Phys.~Rev.~B~\textbf{70},195113 (2004); J. Sun, Y. Zhao, and W. Z. Liang, {\it ibid.}~\textbf{79}, 155112 (2009).





\bibitem{Amico} L. Amico, R. Fazio, A. Osterloh, and V. Vedral. Rev. Mod. Phys.  \textbf{80},517 (2008).





\bibitem{Bera} S. Bera, S. Florens, H. U. Baranger, N. Roch, A. Nazir, and A. W. Chin, Phys. Rev. B.  \textbf{89},121108(R) (2014).





\bibitem{Sil}  R. Silbey and R.A. Harris, J. Chem. Phys.  \textbf{80}, 2615 (1984).





\bibitem{zhao97} Y. Zhao, D. W. Brown, and K. Lindenberg, J.~Chem.~Phys.~\textbf{106}, 2728; \textbf{106}, 5622; \textbf{107}, 3159; \textbf{107}, 3179 (1997).





\bibitem{zhao92} Y. Zhao and H. N. Bertram, J. Magn. Magn. Mater. \textbf{114}, 329 (1992).





\bibitem{zhao95} Y. Zhao and H. N. Bertram, J. Appl. Phys. \textbf{77}, 6411 (1995).





\bibitem{Naz} A. Nazir, D. P. S. McCutheon and A. W. Chin, Phys. Rev. B. \textbf{85}, 224301 (2012).





\bibitem{Lan}  D. P. Landau and K. Binder, \textsl{A Guide to Monte Carlo Simulations in Statistical Physics}, (Cambridge University Press, Cambridge, England, 2005).





\bibitem{Wilson} K. G. Wilson, Rev. Mod. Phys. \textbf{47}, 773 (1975).





\bibitem{bind} K. Binder and D. P. Landau, Phys. Rev. B. \textbf{30}, 1477 (1984)




\bibitem{Dic}  R. Dickman, J. Chem. Phys. \textbf{136}, 174105 (2012).





\bibitem{Sin} I. Sinha and A. K. Mukherjee, J. Stat. Phys. \textbf{146}, 669 (2012).





\bibitem{mps_c} J. I. Cirac and F. Verstraete,  J. Phys. A: Math. Theor. \textbf{42} 504004 (2009); U. Schollwock, Ann. Phys. (Leipzig) \textbf{326}, 96 (2011).



\bibitem{obb} C. Zhang, E. Jeckelmann, and S. R. White, Phys. Rev. Lett. \textbf{80}, 2661 (1998).



\end{thebibliography}
\end{document}